\documentclass[journal,12pt,onecolumn,draftclsnofoot,]{IEEEtran}
\IEEEoverridecommandlockouts
\usepackage{cite}
\usepackage{amsmath,amssymb,amsfonts}
\usepackage{algorithmic}
\usepackage{graphicx}
\usepackage{subcaption}
\usepackage{amsmath}
\usepackage{textcomp}
\usepackage{xcolor}
\usepackage{url}
\usepackage{multirow}
\usepackage{hyperref}
\usepackage[numbers,compress]{natbib}
\usepackage{wrapfig}
\def\BibTeX{{\rm B\kern-.05em{\sc i\kern-.025em b}\kern-.08em
   T\kern-.1667em\lower.7ex\hbox{E}\kern-.125emX}}
\begin{document}

\title{%Assessing the Effectiveness of Single-phase Immersion Cooling for Thermal Management of Li-Ion Batteries using Multi-Physics Numerical Simulations
Multi-Physics Numerical Analysis of Single-phase Immersion Cooling for Thermal Management of Li-Ion Batteries}
\author{%%%% author names
    \IEEEauthorblockN{Piyush Mani Tripathi}% first author
    ~and \IEEEauthorblockN{Amy M. Marconnet}% delete this line if not needed
    
    % duplicate the line above as many times as needed to list all authors
    %%%% author affiliations
    \IEEEauthorblockA{\textit{School of Mechanical Engineering and the Birck Nanotechnology Center, Purdue University, West Lafayette, United States of America}}% first affiliation
    
    % duplicate the line above as many times as needed to list all affiliations
    %%%% corresponding author contact details
    %\IEEEauthorblockA{tripat35@purdue.edu}
}
\maketitle

\begin{abstract}
 Battery thermal management systems (BTMSs) are critical for efficient and safe operation of lithium-ion batteries (LIBs), especially for fast charging/discharging applications that generate significant heating within the cell. Forced immersion cooling, where a dielectric fluid flows in direct contact with the LIB cells, is an effective cooling approach. But because of its complex nature, a thorough understanding of the underlying physics - including the coupled electrochemical, thermal, fluid, and mechanical effects - is required before immersion cooling will see wide adoption into commercial systems. Past research on immersion cooling focused mainly on the coupled thermal-fluid transport problem. But for a holistic assessment of immersion cooled LIB systems, electrochemical and thermal-fluid aspect should be considered simultaneously as both are interdependent.
 In this work, to investigate the performance of a LIB subjected to forced immersion cooling, we develop a fully coupled modeling approach that solves the detailed electrochemical model in conjunction with the thermal-fluid transport models for both the cell and fluid domain. After calculating the electrochemical and thermal responses, we also estimate the mechanical stresses within the cell generated due to the ion diffusion and temperature rise that impact reliability.
 To assess the effectiveness of forced immersion cooling, we evaluate several different configurations for a cylindrical 18650 battery cell (with Nickel Manganese Cobalt (NMC) cathode material) under varying cell discharge rates (1C, 3C \& 5C). We compare forced immersion cooling for two liquids (deionized water and mineral oil) at three different fluid mass flow rates (0.0025, 0.0050 \& 0.0100 kg/s). The liquid cooling results are compared to forced convection cooling with air (at a velocity of 0.3 m/s). 
 As the electrochemical kinetics improve with increasing temperature, the thermal management goals for stability and reliability of the cell compete with the electrochemical performance goals. Specifically, with improved temperature control (lower temperatures), there is a higher capacity loss for a cell, but the lower temperatures and lower temperature variations during operation are expected to improve reliability. This highlights the strong cross-coupling of the electrochemical and heat transfer phenomena.
 Furthermore, a sensitivity analysis of dielectric fluid properties (\textit{i.e.}, density, thermal conductivity, specific heat, and viscosity) on the immersion cooling performance at a fixed mass flow rate and cell discharge rate suggest that the thermal conductivity and the specific heat capacity of the fluid have the most significant effects on temperature rise of the battery.
 By comparing results across fluids and flow rates, we define a new metric that can be used to compare the cooling capacity considering different flow parameters. Overall, this study provides insights that will be useful in the design of immersion cooling-based BTMSs including, for example, the selection of forced immersion cooling specifications (\textit{i.e.}, dielectric fluid and mass flow rate), such that the temperature is controlled without significant capacity loss. The newly defined metric for assessing performance accelerates system design by enabling physics-informed choices without computationally-expensive numerical simulations.

\end{abstract}

\begin{IEEEkeywords}
Thermal Management, Li-Ion battery, Immersion cooling, Dielectric fluid

\end{IEEEkeywords}

\section{Introduction}\label{Introduction}
Recently, the noticeable effects of climate change such as global warming, irregular weather patterns, and higher levels of air pollution have forced people to work towards sustainable growth without exploiting or decimating natural resources. One of the key strategies to achieve this is to reduce our reliance on non-renewable energy sources such as fossil fuels (\textit{i.e.}, oil, coal) in all areas \cite{Chu2012OpportunitiesFuture}. As a result, electricity generation from renewable sources (such as solar, wind, and geothermal) is increasing significantly. Because of the transient nature of some of these sources (e.g., wind and solar) and the need for portable power in the transport sector, energy storage devices including batteries are now indispensable. For example, electric vehicles (EVs) ranging from cars to trucks are now ubiquitous on the roads. Lithium-ion batteries (LIBs) are one of the prominent energy storage technologies because of their high energy density, high power density, and other secondary considerations such as cost and lifetime \cite{Wang2016AVehicles, Dunn2011ElectricalChoices}.

However, there are inherent thermal challenges that hinder efficient and safe functioning of LIBs. For example, fast charging/discharging rates lead to high operating temperatures that accelerate the degradation (aging) of the cells and increase the chances of catastrophic failure (thermal runaway). Additionally, lower working temperatures reduce the discharge capacity \cite{Wang2016AVehicles, Zeng2021ACharging}, but can improve stability and lifetime. Previous studies have demonstrated that there is an optimum temperature range to circumvent the effects of extreme temperatures (\textit{i.e.}, high \textit{and} low temperatures) in order to achieve the best performance for a lithium-ion battery cell \cite{Hao2018EfficientAlloy}. Typically, the desired operational temperature (across a wide range of charging/discharging rates) is between 20$^{\circ}$C and 40$^{\circ}$C for efficient performance with a long lifespan \cite{Lin2021ASystem}. Given the internal heat generation during normal operation and varying environmental conditions, an effective battery thermal management system (BTMS) becomes an integral part of a LIB system.

Battery thermal management systems (BTMSs) can be broadly categorized as active or passive. \textit{Active} systems use forced fluid (liquid or gas) flow to remove heat, whereas \textit{passive} systems store the heat in some media (such as phase change materials (PCMs) that store energy through melting) and release it at a later time (for PCMs, when re-solidifying) \cite{KumarThakur2023AFast-charging}. One of the most effective active cooling mechanisms is immersion cooling \cite{Roe2022ImmersionReview}, where a fluid (usually a dielectric fluids to ensure electric insulation) comes in direct contact with the LIB cells to moves the heat out of the battery pack or module. There are primarily two categories of immersion cooling systems: \textit{static immersion cooling} \cite{Liu2023Single-phaseModule, Jilte2019CoolingSystems} (where the LIBs are submerged in a static pool of dielectric fluid and fluid motion is due primarily to buoyancy effects) and \textit{forced immersion cooling} \cite{Huo2015TheMethod, Liu2023NumericalCooling, Jithin2022NumericalFluids} (where a dielectric fluid is circulated through the battery pack or module around each cell). From a heat transfer perspective, forced immersion cooling is expected to perform better than static immersion cooling because of the enhanced fluid flow rates, which result in higher cooling capacity, and the FIC should be less sensitive to driving conditions for EV applications.

 Although forced immersion cooling is a promising candidate for an effective thermal management system, the design of these sophisticated systems requires a thorough understanding of the strong cross-coupling of the electrochemical, heat transfer, fluid flow, and mechanical effects. Previous studies have focused primarily on the thermal-fluid aspect with simple models  (\textit{e.g}, similar to Ohm's Law of heat generation \cite{Huo2015TheMethod, Liu2023NumericalCooling, Solai2022ValidationPack} or semi-empirical models \cite{Jithin2022NumericalFluids}) to account for the electrochemistry that governs the rate of heat generation. Also, numerical work on immersion cooling has generally been restricted to low discharge rates ($\leq$ 3C), which correspondingly have lower heat generation rates. The findings may not be relevant for fast-charging applications. Moreover, at higher discharge rates, the non-linearity becomes more prominent since temperature significantly influences the LIB electrochemical performance as has been demonstrated by studies \cite{Lyu2020InvestigationStorage, Dong2018NumericalOperations} that vary the ambient temperature and/or the convection heat transfer coefficients. This multiphysics coupling within the system demonstrates the need for a detailed and holistic approach to understanding the performance of immersion-cooled battery systems considering the coupling of temperature, electrochemistry, and heat generation. Typically, fully coupled multi-physics models are computationally expensive. Thus, for efficient design of such systems, rapid design estimates using physics-informed performance metrics to analytically compare different cooling solutions would accelerate the design process.   

 In the present study, we first develop a comprehensive numerical model that solves a detailed electrochemical model in conjunction with a 3D thermal-fluid model to investigate the impact of single-phase forced immersion cooling on the electrochemical performance and the thermal response. After calculating the electrochemical and thermofluid response, the potential for mechanical degradation is assessed by examining the resulting diffusion and thermal stresses. To get a thorough insight into the optimal design parameters for immersion cooling, different combinations of discharge rates, mass flow rates, and dielectric fluids are evaluated. Furthermore, the impact of the properties of the dielectric fluids (\textit{e.g.}, density, thermal conductivity, specific heat, and viscosity) on the battery system performance are evaluated through a detailed set of parametric sweeps. From these results, we identify and demonstrate the efficacy of using a newly proposed metric for analyzing the relative performance of forced immersion cooling configurations. Note that during charging/discharging of LIB cell same order of heat is generated \cite{Zhao2018ThermalFlow} since the physical process is very similar (\textit{i.e.}, lithiation$\backslash$delithiation of the electrodes) and usually the temperature rise is even slightly higher for the discharging process compared to charging operation for a given cooling system \cite{Dong2018NumericalOperations, Prada2012Applications}. For this reason, present study focuses on discharging process, however the findings of this study will be valid for immersion cooling of LIB cell irrespective of charging$\backslash$discharging operation since the underlying physics remain the same.

\section{Methodology}
In this section, we first describe the details of the forced immersion cooling numerical model including the geometry (the LIB cell and domain for fluid flow), details of each sub-model (\textit{i.e}, electrochemical, fluid-thermal \& mechanical), and the parameters including material properties.

\begin{figure*}
    \centering
    \includegraphics[scale=0.6]{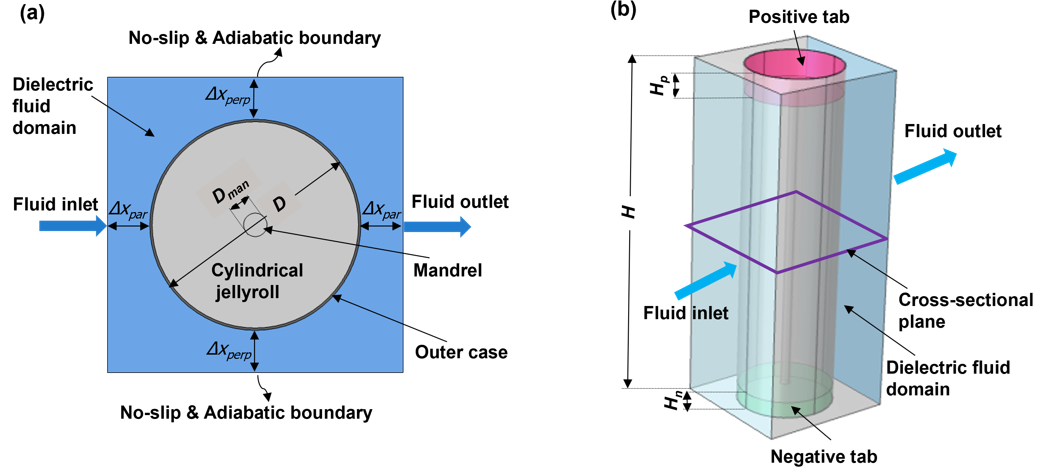}
    \caption{(a) Cross-sectional and (b) 3D schematics of the immersion cooling geometry including a single 18650 battery cell and the neighboring fluid domain. All the key dimension along with the cross-section plane is highlighted.}
    \label{fig: Geometry schematic comb}
\end{figure*}

% \begin{figure*}
%      \centering
%      \begin{subfigure}[b]{0.49\textwidth}
%          \centering
%          \includegraphics[width=\textwidth]{geom_cross section.png}
%          %\caption{}
%          \label{Fig. 1(a)}
%      \end{subfigure}
%      \hfill
%      \begin{subfigure}[b]{0.35\textwidth}
%          \centering
%          \includegraphics[width=\textwidth]{Battery Model-1/geom_3D_final.png}
%          %\caption{}
%          \label{Fig. 1(b)}
%      \end{subfigure}
%      \hfill
%      % \begin{subfigure}[b]{0.6\textwidth}
%      %     \centering
%      %     \includegraphics[width=\textwidth]{Battery Model-1/algorithm.png}
%      %     %\caption{}
%      %     \label{Fig. 1(c)}
%      % \end{subfigure}
%      % \hfill
%         \caption{(a) Cross-sectional and  (b) 3D schematics of the immersion cooling geometry including a single 18650 battery cell and the neighboring fluid domain.  %(c) Flowchart detailing the algorithm used to simulated the electrochemical, thermo-fluid, and mechanical response of the battery system.
%         }
%         
% \end{figure*}

\subsection{Simulated Geometry}\label{Geometry}
 The forced immersion cooling geometry includes a single cylindrical LIB cell and the surrounding fluid domain (see Figure \ref{fig: Geometry schematic comb}(a-b). The battery cell itself includes a cylindrical ``jellyroll'' (consisting of alternating layers of anode, cathode, and separator materials soaked with a liquid electrolyte) surrounding a central mandrel and enclosed in an outer case \cite{Quinn2018EnergyCells, Jeon2011ThermalCycle}. Electrical connections are made at the top of the bottom of the cell where the current collectors for the cathodes and anodes inside the cell are connected to the outside of the case. 
 
In the present study, an 18650 Nickel Manganese Cobalt (NCM)-based cylindrical LIB cell is selected for analysis, since NMC has one of the highest energy densities for LIB batteries \cite{Wang2016AVehicles}. 
Note that 18650 means that the cell diameter, $D$, is 18 mm and the height, $H$, is 65 mm.  Chemical reactions, including those that generate heat, primarily occur within the jellyroll.  Key geometrical parameters of the cylindrical cell include the mandrel diameter ($D_{man} \approx$ 2~mm), the outer case thickness ($\approx$~0.2~mm), the thickness of positive tab ($H_p \approx$~5~mm), and the thickness of the negative tab ($H_n \approx$~3 mm), which are estimated based from literature \cite{Kang2022HowBatteries, Quinn2018EnergyCells}.
 
  %The cylindrical jellyroll (which is a long sheet of cathode, anode \& separator pressed together and soaked with electrolyte) wrapped around a mandrel encased in a outer can which is sealed with a positive tab on top and a negative tab on the bottom as shown in Figure \ref{fig: Geometry schematic}.  
 % \cite{Quinn2018EnergyCells, Jeon2011ThermalCycle}
  
 The dielectric fluid domain is a cuboidal region around the cylindrical cell (see Figure \ref{fig: Geometry schematic comb} (b)). The fluid domain completely surrounds the lateral curved surface of the cell and the height of the flow domain is selected to be the same as the cell height, $H$. Note that the dielectric fluid does not flow over the top of positive tab or below the negative tab for this analysis.\footnote{Note that the height of the fluid domain can also be taller than the cell height $H$, which adds cooling directly to the top electrode (\textit{i.e.}, adding `tab cooling'). Results when including tab cooling are briefly discussed in \hyperref[TabCooling]{Appendix B}.} 
 The cross-section of flow domain, which governs the flow dynamics around the cell, is controlled by two key geometrical parameters: $\Delta x_{par}$ and $\Delta x_{perp}$, which are the minimum distances from the case surface to the wall in directions parallel (subscript `par') and perpendicular (subscript `perp') to the flow. In the present study, $\Delta x_{perp}$ and $\Delta x_{par}$ are both chosen to be $0.2 D$. Thus, the total volume of the simulated domain is $(D+2\Delta x_{par}) \times (D+2\Delta x_{perp}) \times H$ or $1.4 D \times 1.4 D \times H$.

 During operation (\textit{i.e.}, charge\textbackslash discharge) of a LIB cell, electrochemical reactions occur within the active materials that make up the jellyroll. As the LIB operates, heat generated by these processes must diffuse (conduct) through the components that make up the cell and then dissipate by convection into the dielectric fluid at the surface of the cell. The cross-coupled nature of physics for electrochemistry, heat transfer, and fluid flow makes it challenging to estimate performance accurately without computationally-intensive numerical simulations. 

\begin{table}[t]
    \caption{Governing Equations for the Pseudo-2D (P2D) Model of the Battery Cell }
    \begin{center}
    \begin{tabular} { |p{3.8cm}|p{2.1cm}|p{8.8cm}| }
         \hline
         \textbf{Description} & \textbf{Domain} & \textbf{Governing equation} \\
          \hline
         Conservation of Li\textsuperscript{+} Species & Electrode & $\frac{\partial c_{s}}{\partial t}= \frac{1}{r^{2}}\frac{\partial }{\partial r}\left ( D_{s}r^{2}\frac{\partial c_{s}}{\partial r} \right )$  \\
          \hline
          Conservation of Li\textsuperscript{+} Species  & Electrolyte & $\varepsilon \frac{\partial c_{e}}{\partial t}= \triangledown.\left ( D_{e}^{eff}\triangledown c_{e} \right)+\left ( 1-t_{+} \right )aj$ \\
          \hline
          Charge Conservation & Electrode & $\triangledown.\left ( \sigma_{s}^{eff}\triangledown\phi_{s} \right )- aFj = 0$ \\
           \hline
          Charge Conservation &  Electrolyte & $\triangledown.\left ( \kappa^{eff}\triangledown\phi_{e} \right )+ \triangledown.\left ( \kappa_{D}^{eff}\triangledown \ln{c_{e}} \right ) + aFj = 0$ \\
           \hline
         Electrochemical Kinetics & Electrode-Electrolyte interface & $j= i_{0}\left \{ exp\left ( \frac{0.5F}{RT}\left (  \phi_{s}- \phi_{e}-U\right ) \right )-exp\left ( -\frac{0.5F}{RT}\left (  \phi_{s}- \phi_{e}-U\right ) \right )\right \}$ \\
           \hline 
    \end{tabular} \label{Tab:P2D model governing equation}
    \end{center}
\end{table}
\subsection{Coupled Numerical Model for the Immersion Cooling Battery System}

% Figure \ref{fig: Geometry schematic}(c) shows an overview of the algorithm for solving the coupled electrochemical, thermo-fluid, and mechanical response of the system.
The following sub-sections discuss the modeling approach used here to understand the impact of immersion cooling on performance from electrochemical, thermal, and mechanics/reliability perspectives.

 As mentioned in Section \ref{Geometry}, a LIB consists of a cylindrical jellyroll (which consists of a long sheet of cathode, anode, and separator pressed together wrapped around a mandrel) encased in an outer can that is sealed with a positive tab on top and a negative tab on the bottom. This geometry contains features across different length scales and integrates numerous thin layers of different materials with large aspect ratios. Thus, simplified models are needed for accurate, but efficient, analysis of the system. 
 
 Here, we develop a psuedo-two dimensional (P2D) electrochemical model (that models a single anode, cathode, and separator in order to represent the entire jellyroll domain), while the heat conduction and thermo-fluid models are fully three-dimensional. The simulation domain of the P2D model (\textit{i.e.}, the anode, separator, and cathode) is the building block of LIB and is arranged to achieve the geometry and performance of the cell.  He \textit{et al.} \cite{He2022AProcess} demonstrated that assuming the cylindrical jellyroll to be a homogeneous region for the electrochemical model (instead of separate domains for the individual anode, separator, and cathode in 3D) does not significantly affect the accuracy of thermal analysis of a LIB.

\subsubsection{\textbf{Electrochemical Model of the Battery Cell}}\label{P2D_model}
In the present study, the electrochemical response of the battery cell is modeled following a Neumann pseudo-two dimensional (P2D) approach  \cite{Doyle1993ModelingCell, Fuller1994SimulationCell}. The P2D model approximates the 3D cell as 1 anode-separator-cathode zone based on the total active electrode area.  The simulation domain consists of a porous anode (Li\textsubscript{x}C\textsubscript{6}, Graphite), a porous separator, and a porous cathode (LiNi\textsubscript{1/3}Co\textsubscript{1/3}Mn\textsubscript{1/3}O\textsubscript{2} or `NMC' for short). This domain is impregnated with a liquid electrolyte, which is a lithiated organic solution. During charging$\backslash$discharging of a cell, diffusion of Li\textsuperscript{+} ion occurs between two electrodes (\textit{i.e.}, cathode \& anode) via electrolyte as well as within the electrodes, which eventually changes the electric potential of the cell. This phenomenon is simulated by conserving charge and species in the cell domain. All the equations except the diffusion of Li\textsuperscript{+} are solved in 1D along the thickness of the cell, whereas the diffusion equation in electrodes is solved in the radial direction of spherical coordinates as the active particles in electrode are assumed to be spherical. 

\begin{table}[t]
    \caption{Parameters of the P2D Electro-Chemical Model}
    \begin{center}
    \begin{tabular} { |p{8cm}|c|c|c|c|}
         \hline
         \multicolumn{5}{|c|}{\textbf{Geometric Details}} \\
         \hline
         \textbf{Parameter} & \textbf{Anode} & \textbf{Separator} & \textbf{Cathode} & \textbf{Ref}\\
          \hline
         Layer Thickness, $l$ ($\mu$m) &  80  & 25 & 73 & \cite{Kang2022HowBatteries}\\
          \hline
         Layer Porosity, $\epsilon$ (-)  &  0.3  & 0.4 & 0.3 & \cite{Kang2022HowBatteries} \\
          \hline
         Particle Radius, $r_p$ ($\mu$m) & 5 & N.A. & 5 & \cite{Kang2022HowBatteries}\\
           \hline
        Volume Fraction of Active Particles, $\epsilon_a$ (-) &   0.662 & N.A. & 0.58 & \cite{Kang2022HowBatteries}\\
           \hline  
        Specific Surface Area, $a$ (1/m) & \multicolumn{3}{|c|}{3$\epsilon/r_p$} & \cite{Kang2022HowBatteries}\\
           \hline 
        Bruggermann Exponent, $\beta$ & \multicolumn{3}{|c|}{1.5} & \cite{Kang2022HowBatteries}\\
           \hline 
        Electronic Conductivity, $\sigma_s$ (S/m) &   10 & N.A. & 0.1 & \cite{Kang2022HowBatteries}\\
           \hline 
        Reference Reaction Rate, $k_{ref}$ (m$^{2.5}$mol$^{-0.5}$s$^{-1}$) &   2.32$\times 10^{-10}$ & N.A. & 2.4177$\times 10^{-11}$ & \cite{Chen2017ProbingBehavior},[est$^1$]\\
           \hline 
        Reference Diffusion Coefficient, $D_s$ (m$^{2}$s$^{-1}$) & 1.4$\times 10^{-14}$ &  & 2$\times 10^{-14}$ & \cite{Kang2022HowBatteries}\\
           \hline 
        Transference Number, $t^+$ & \multicolumn{3}{|c|}{0.363} & \cite{Kang2022HowBatteries, Chen2017ProbingBehavior}\\
           \hline 
        Initial Li\textsuperscript{+} Concentration in Electrolyte, $c_e^0$ (mol/m$^3$) & \multicolumn{3}{|c|}{1200} & \cite{Kang2022HowBatteries, Ji2013Li-IonTemperatures}\\
           \hline 
       Maximum Li\textsuperscript{+} Concentration in the Electrodes, $c_s^{max}$ (mol/m$^3$) & 30,900 & N.A. & 49,500 & \cite{Chen2017ProbingBehavior}\\
           \hline 
    Activation Energy for $D_s$ ($E_{act,d}$) (kJ/mol) & 30 & N.A. & 30 & \cite{Ji2013Li-IonTemperatures}\\
           \hline 
    Activation Energy for $k$  ($E_{act,k}$) (kJ/mol) & 68 & N.A. & 50 & \cite{Ji2013Li-IonTemperatures}\\
           \hline 
     Contact Resistance, $R_c$ ($\Omega$~cm$^2$) & \multicolumn{3}{|c|}{6} & \cite{Ji2013Li-IonTemperatures}\\
           \hline 
     Electrode Area, $A$ (cm$^2$) & \multicolumn{3}{|c|}{600} & \cite{Ji2013Li-IonTemperatures}\\
           \hline 
    \multicolumn{5}{c}{$^1$Note that `est' means that parameter is estimated by comparing numerical and experimental results.}
    \end{tabular} \label{Tab::P2D model parameters}
    \end{center}
   % \footnote{Note that `est' means that parameter is estimated by comparing numerical and experimental results.}
\end{table}

Table \ref{Tab:P2D model governing equation} summarizes the governing equations for the P2D electrochemical model. Note that thermal effects are coupled to the electrochemical response through the temperature-dependent properties and heat generation rates using an approach similar to past research \cite{Gu2000Thermal-ElectrochemicalSystems}. The thermal model is described in detail in Section \ref{sec:thermalModel} and provides spatial- and time-varying temperatures within the battery cell. All parameters and properties used in the P2D simulation are outlined in Table \ref{Tab::P2D model parameters} along with the references from which it has been adopted. Note that throughout the simulations all the variables are estimated in appropriate SI units as described in the table. Key variables of interest in P2D model are the concentration of Li\textsuperscript{+} species in the solid phase of the porous electrodes ($c_s$) and the electrolyte ($c_e$), the electric potential in solid phase of porous electrodes ($\phi_s$) and the electrolyte ($\phi_e$), and the transfer current density ($j$). 

The transfer current density governs the electrochemical kinetics \cite{Cai2009ReductionSimulations,Chen2017ProbingBehavior,Cai2011MathematicalSoftware} at the electrode-electrolyte interface and are defined by the Butler-Volmer equation (see Table~\ref{Tab:P2D model governing equation}). The transfer current density depends on the exchange current density ($i_0$), which is a function of the Lithium concentration in electrodes and electrolyte:
\begin{equation}\label{i0}
 i_0 = k(T)(c_s^{max}-c_{s,surf})^{0.5}c_{s,surf}^{0.5}c_e^{0.5}
\end{equation}
where $c_s^{max}$ is the maximum possible concentration of Li\textsuperscript{+} in each electrodes (\textit{i.e.}, cathode \& anode), $c_{s,surf}$ is the Li\textsuperscript{+} concentration at the surface of electrode particles, and $k$ is the temperature dependent reaction rate for each electrode. Note that the transport and kinetic properties of anode and cathode in the P2D model are temperature-dependent, and their values are calculated using an Arrhenius-type equation:
\begin{equation}\label{Arhenius eqn}
 \varphi =\varphi _{ref}\left ( \frac{E_{act}}{R}\left ( \frac{1}{T_{ref}}-\frac{1}{T} \right ) \right )
\end{equation}
where $\varphi$ is the temperature-adjusted property (such as $k$, $D_s$, ...) as a function of temperature ($T$), $\varphi_{ref}$ is the value of the property at the reference temperature ($T_{ref}$), and $E_{act}$ is the activation energy. For the present study, $T_{ref}$ is set at 298~K, which is the same as the initial temperature ($T_i$). For the P2D approach, the temperature $T$ corresponds to the \textit{average} temperature of the entire jellyroll domain calculated from the thermal model (see Eq. \ref{thermal model of cell}).

In the present study, electrolyte properties, such as the ionic conductivity ($\kappa$) and the diffusion coefficient ($D_e$), are concentration- and temperature-dependent \cite{Cai2011MathematicalSoftware,Kumaresan2008ThermalCell,Chen2017ProbingBehavior}:
\begin{multline}\label{electrolyte conductivity}
    \kappa = (10^{-4})c_e\times(-10.5 +0.668\times0.001\times c_e + 0.494\times10^{-6}\times c_e^2\\ +0.074T - 1.78\times10^{-5}c_eT-8.86\times10^{-10}c_e^2T -6.96\times10^{-5}T^2 + 2.8\times10^{-8}c_eT^2)^2 
\end{multline}
and
\begin{equation}\label{electrolyte diffusion}
D_e = 10^{-4}\times10^{(-4.43-54/(T-229-5\times10^{-3}c_e) -0.22\times10^{-3}c_e)}.
\end{equation}
Effective ionic and transport properties of electrodes and electrolytes (\textit{i.e.}, $\sigma_{s}^{eff} = \sigma_s\epsilon_a^\beta$, $\kappa^{eff}=\kappa\epsilon^\beta$ and $D_e^{eff} =D_e\epsilon^\beta$) are determined from the porosity ($\epsilon$) using the Bruggermann exponent ($\beta$), similar to past studies \cite{Cai2011MathematicalSoftware,Cai2009ReductionSimulations,Kang2022HowBatteries}. Additionally, the effective diffusion ionic conductivity ($\kappa_d^{eff}$) depends on $\kappa^{eff}$ and the thermodynamic factor ($V$), which is a function of $c_e$ and $T$ as given by \cite{Chen2017ProbingBehavior,Kang2022HowBatteries}:
\begin{equation}\label{electrolyte diffusion ionic}
\kappa_d^{eff} = \kappa^{eff}V(c_e,T),~~V = 0.601 -0.24(0.001c_e)^{0.5} +0.982(1-0.0052(T - 294))(10^{-9}c_e^3)^{0.5}.
\end{equation}

Additionally, note that the open circuit potential ($U$) and the entropic heating coefficient $\left( \frac{\partial U}{\partial T} \right)$ are functions of parameters of the LIB cell components (\textit{i.e.}, the NMC and graphite electrodes in this study) \cite{Kang2022HowBatteries, Ji2013Li-IonTemperatures,Chen2017ProbingBehavior}. Other constants needed for the P2D model include the Faraday constant ($F$ = 96,485 C/mol) and the universal gas constant ($R$ = 8.3143 J/(mol~K)).

\subsubsection{\textbf{Thermal Model of the Battery Cell}}\label{sec:thermalModel}
A numerical, transient, 3D heat conduction model estimates the temperature response of the cylindrical cell throughout the jellyroll, outer case including the positive and negative tabs, and the mandrel in response to the heat generation calculated based on results from the electrochemical side \cite{Dong2018NumericalOperations, Lyu2020InvestigationStorage, Ji2013Li-IonTemperatures, Chen2017ProbingBehavior, Cai2011MathematicalSoftware, Cai2009ReductionSimulations} (see Section~\ref{P2D_model}) and the convection cooling provided from the thermofluid model (see Section \ref{fluid_flow}). All domains within the conduction model are assumed to be in perfect thermal contact, such that there is continuity of temperature across the domains. Note that the heat generation calculated from the P2D electrochemical model (assuming the average cell temperature at a particular time step) is assumed to provide a uniform volumetric heat source within the homogeneous jellyroll region, which is the active part of the LIB cell, in the thermal model. Thermophysical parameters used as inputs to the numerical model are given in Table~\ref{Thermo_Properties}. The mandrel domain is usually considered to be a void \cite{Quinn2018EnergyCells} and, for the present study, is assumed to be filled with air. Note that apart from the jellyroll, all materials are assumed to have isotropic thermal conductivity ($\lambda$). The anisotropic thermal conductivity of the jellyroll is modeled with a high thermal conductivity in the axial direction ($\lambda_z$ along the sheets of the current collectors) and a lower thermal conductivity in the radial and azimuthal directions ($\lambda_r$ and $\lambda_\theta$). 

\begin{table}[tbh!]
    \caption{Thermophysical Properties of the Modelled LIB Cell.}\label{Thermo_Properties}
    \begin{center}
    \begin{tabular} { |p{3.5cm}|c|c|c|c|}
         \hline
         \multicolumn{5}{|c|}{\textbf{Thermophysical Properties}} \\
         \hline
         \textbf{Region} & \textbf{$\rho$ (kg/m$^3$)} & \textbf{$C_p$ (J/(kg~K))} & \textbf{$\lambda$ (W/(m~K))} & \textbf{Ref.}\\
          \hline
         Jellyroll &  2055  & 1399 & $\lambda_r = \lambda_{\theta} = 1.32$; $\lambda_z = 19.62$& \cite{Lyu2020InvestigationStorage}\\
          \hline
         Outer Case  &  7500  & 460 & 14 & \cite{Cheng2019InfluencesBattery, Li2021OptimalBatteries} \\
          \hline
         Positive Tab & 2719 & 871 & 202 & \cite{Li2021OptimalBatteries,Jithin2022NumericalFluids}\\
           \hline
         Negative Tab & 8978 & 381 & 387 & \cite{Li2021OptimalBatteries,Jithin2022NumericalFluids}\\
           \hline 
    \end{tabular}
    \end{center}
\end{table}

The local transient temperature of the battery cell is determined for the 3D cylindrical cell domain from the heat diffusion equation (\textit{i.e.}, conservation of energy) applied within each component of the system:
\begin{equation}\label{thermal model of cell}
 \rho C_{p} \frac{\partial T(r,\theta,z,t)}{\partial t}= \frac{1}{r}\frac{\partial }{\partial r}\left ( r\lambda_r\frac{\partial T}{\partial r}  \right )+ \frac{1}{r^2}\frac{\partial }{\partial r}\left ( \lambda_{\theta}\frac{\partial T}{\partial \theta}  \right ) + \frac{\partial }{\partial z}\left ( r\lambda_z\frac{\partial T}{\partial z}  \right ) %\lambda\triangledown^2T(x,y,z,t) 
 +\dot{q}(t)
\end{equation}
where $t$ is time, $\rho$ and $C_p$ are the density and specific heat of each domain simulated, and $\dot{q}$ is the volumetric heat generation. %In the present study, the aniostropy of the jellyroll domain is accounted using different values of radial, $\lambda_r$ ($=\lambda_{\theta}$) and axial, $\lambda_z$ thermal conductivity. For all other domains, thermal conductivity is  isotropic, $\lambda=\lambda_r=\lambda_{\theta} =\lambda_z$ and therefore represented by a single value.

The volumetric heat generation is calculated based on the output of P2D model and is only nonzero in the cylindrical jellyroll region. Specifically, it can be determined as: 
\begin{multline}\label{heat generation}
   \dot{q} = \underbrace{j(\phi_{s}- \phi_{e}-U)}_{\text{kinetic heating},~\dot{q}_{rxn}} + \underbrace{ j T\left ( \frac{\partial U}{\partial T} \right )}_{\text{reversible heating},~\dot{q}_{rev}} + \underbrace{\sigma_{s}^{eff} \triangledown \phi_{s}\cdot\triangledown\phi_{s} + \kappa^{eff} \triangledown \phi_{e}\cdot\triangledown\phi_{e} + \kappa_{D}^{eff} \triangledown \ln{c_{e}}\cdot\triangledown\phi_{e}}_{\text{Ohmic heating},~\dot{q}_{ohm}} 
\end{multline}
The first set of terms $\left( \dot{q}_{rxn}=j(\phi_{s}- \phi_{e}-U) \right)$ is the kinetic heating; the second term $\left({\dot{q}_{rev}} = jT\left ( \frac{\partial U}{\partial T} \right) \right)$ is the reversible heat source; and all other terms contributes to Ohmic heating (${\dot{q}_{ohm}}$), which is a result of electronic resistance, ionic resistance, and concentration overpotential. Apart from the electrochemical heat generation term given in $\dot{q}$ above, contact resistance results in heat generation during cell operation, $\dot{q}_{cont} \sim I^2\frac{R_c}{A} $, where $I$ is the current and $\dot{q}_{cont}$ has been accounted in the thermal model.

Initially, the cell (and fluid domain) are assumed to be at a uniform temperature of $T_i$ = 298.15~K. Heat generated within the cell leads to the cell temperature to rise and the heat is dissipated by convection to the fluid domain. Surface energy balances at the interface between the solid and liquid domains link the conduction within the cell (related to the temperatures and temperature gradients in these materials) to heat transfer within the fluid. 

%Thermal properties of the all the five domains of the cylindrical cell (\textit{i.e.}, jellyroll, outer can, mandrel, positive tab \& negative tab) simulated in the thermal model are mentioned in Table \ref{Tab::Thermal and mechanical properties}.

\subsubsection{\textbf{Thermo-fluid Model within the Dielectric Fluid Domain}}\label{fluid_flow}

For the understanding the coupled fluid flow and heat transfer, conservation equations are solved within the the domain of the dielectric fluid, while insuring conservation of energy at the solid-liquid interface.

The governing equations are conservation of mass, momentum, and (thermal) energy within in the 3D fluid domain:
\begin{equation}\label{Continuity eqn}
\frac{\partial \rho_f}{\partial t}+\triangledown \cdot \left ( \rho_{f} \vec{v}\right )=0,
\end{equation}
\begin{equation}\label{Momentum eqn}
\frac{\partial \left ( \rho_f \vec{v} \right )}{\partial t}+\triangledown \cdot \left ( \rho_{f} \vec{v}\vec{v} \right )=-\triangledown p+\triangledown\cdot \left ( \mu_f\triangledown \vec{v} \right ),
\end{equation}
and
\begin{equation}\label{Energy Conservation eqn}
\frac{\partial \left ( \rho _fC_{p,f}T  \right )}{\partial t}+\triangledown \cdot \left ( \rho _fC_{p,f} T \vec{v} \right )=\triangledown \cdot \left ( \lambda_f\triangledown T \right ),
\end{equation}
where $\rho_f$ is density of fluid, $\vec{v}$ is velocity, $p$ is pressure, $\mu_f$ is dynamic viscosity, $C_{p,f}$ is the specific heat, and $\lambda_f$ is the thermal conductivity of the fluid. Note that we are neglecting the effects of gravity.

For the fluid flow, the boundary conditions are as follows:
\begin{itemize}
    \item \textit{Inlet Face}: specified mass flow rate (or velocity for air cooling comparison);
    \item \textit{Outlet Face}: specified pressure; and
    \item \textit{Other Boundaries} (including the interface with the case and the other external walls of the fluid domain): No-slip condition.
\end{itemize}
The inlet boundary conditions (\textit{i.e.}, mass flow rates for the liquids and fluid velocity for gasses) are selected such that fluid flow is in the laminar region (although this can be relaxed with appropriate turbulent flow closure relations). 

For the required thermal boundary conditions, the inlet temperature is fixed at the initial temperature ($T_i$ = 25~$^{\circ}$C = 298.15~K). All other exterior walls except the inlet and outlet faces are assumed to be adiabatic. Surface energy balances at the case-fluid interface dictate the heat fluxes on those boundaries. 

Two different dielectric fluids (deionized water (DIW) and mineral oil (MO)) are chosen for the present analysis. Also, forced air cooling is simulated for comparison. Table~\ref{Tab::Dilectric fluid Props} shows the thermophysical properties of both the dielectric fluids and air. Note that the properties of fluids are assumed to be constant in the simulation. 

\begin{table}[tbh!]
    \caption{Thermophysical Properties of the Selected Dielectric Fluids and Air}\label{Tab::Dilectric fluid Props}
    \begin{center}
    \begin{tabular} { |p{5.5cm}|c|c|c|}
         \hline
         \textbf{Properties} & \textbf{De-Ionized Water} & \textbf{Mineral Oil} & \textbf{Air} \\
         \multicolumn{1}{|r|}{Abbreviation:}   & \textbf{DIW}  & \textbf{MO} & \\
         \multicolumn{1}{|r|}{References:}  & \cite{Jithin2022NumericalFluids,Sirikasemsuk2021ThermalModule}  & \cite{Chen2016ComparisonCells,Jithin2022NumericalFluids}& \cite{Chen2016ComparisonCells,Jithin2022NumericalFluids}\\
          \hline
         Density, $\rho_f$  (kg/m$^3$) &   998  & 920 & 1.225\\
          \hline
         Specific Heat Capacity, $C_{p,f}$  (J/(kg K))  & 4182  & 1900 & 1006\\
          \hline
         Thermal Conductivity, $\lambda_f$   (W/(m K)) & 0.6 & 0.13 & 0.0242\\
           \hline
         Dynamic Viscosity, $\mu_f$  (Pa s) &   0.001 & 0.05 & 1.79$\times 10^{-5}$\\
           \hline           
    \end{tabular} 
    \end{center}
\end{table}

\subsubsection{\textbf{Mechanical Model of the Cell}} Mechanical stress is a key factor that influences both performance and degradation of battery cells  \cite{Fu2013ModelingBattery, Xiao2010ABattery, Duan2018ABatteries, Shao2023AExpansion, Li2020FractureMaterials}. Two factors contribute significantly to the mechanical response within the cell: temperature (or thermal strain ($\varepsilon_{ij}^{th}$)) and diffusion (or diffusion strain ($\varepsilon_{ij}^{diff}$)), assuming the outer rigid outer case restricts the expansion of the cell. In the present study, the mechanical model estimates the average stress generated during the operation in the cylindrical jellyroll region, which is the active region of LIB, based on the temperatures and electrochemical state of the cell. Table~\ref{Mech_Properties} shows the mechanical properties of the different layers in the model.

\begin{table}[bth!]
    \caption{Mechanical Properties of the Modelled LIB Cell.}\label{Mech_Properties}
    \begin{center}
    \begin{tabular} { |p{3.5cm}|c|c|c|c|}
         \hline         
          \multicolumn{5}{|c|}{\textbf{Mechanical Properties}} \\
         \hline
         \textbf{Material} & \textbf{$E$} (GPa)& \textbf{$\mu$} & \textbf{$\alpha$} (1/K)& \textbf{Ref.}\\
           \hline
       NMC Electrode & 142 & 0.3 & 8.6$\times$10$^{-6}$ &  \cite{Xu2017Study,Fu2013ModelingBattery,Xiao2010ABattery, Duan2018ABatteries}\\
           \hline 
        Graphite Electrode & 10 & 0.3 & 4$\times$10$^{-6}$ &\cite{Fu2013ModelingBattery,Xiao2010ABattery, Duan2018ABatteries,Shao2023AExpansion}\\
           \hline 
        Separator & 0.14 & 0.3 & - &\cite{Fu2013ModelingBattery,Xiao2010ABattery, Duan2018ABatteries,Shao2023AExpansion}\\
           \hline 
        Positive Current Collector & 70 & 0.3 & 23$\times$10$^{-6}$ & \cite{Fu2013ModelingBattery,Xiao2010ABattery, Duan2018ABatteries}\\
           \hline 
        Negative Current Collector & 110 & 0.3 & 17$\times$10$^{-6}$ &\cite{Fu2013ModelingBattery,Xiao2010ABattery, Duan2018ABatteries}\\
           \hline 
    \end{tabular} 
    \end{center}
\end{table}

\textit{Diffusion-induced stresses} ($\sigma_{diff}$) are a consequence of delithiation or lithiation of the porous electrodes (\textit{i.e.}, the cathode and anode) during the operation of LIB. Specifically, the addition or removal of a species (in this case, Li\textsuperscript{+}) from a material causes a change in the volume of the materials. Typically, the diffusive stress, $\sigma_{diff}$, is estimated based on the diffusion-thermal analogy \cite{Xiao2010ABattery}:
\begin{equation}\label{diffusion strain}
\varepsilon_{ij}^{th} = \frac{1}{3}\Delta c\Omega\delta_{ij}.
\end{equation}
where $\delta_{ij}$ is the Dirac delta function and $\Delta c$ is the change in concentration of diffusion species. 
Here, we calculate the diffusion stress based on analytical expressions proposed in a recent study \cite{Shao2023AExpansion}:
\begin{equation}\label{diffusion stress of cell}
 \sigma_{diff} = -K(\Delta l_{c}^{diff} + \Delta l_{a}^{diff}),
\end{equation}
where $\Delta l_{c}^{diff}$ and $\Delta l_{a}^{diff}$ are the change in the thickness of the cathode and anode due to diffusion (subscripts `c' and `a', respectively). A key assumption in this equation is that ion transport does not cause a change in thickness of the separator $\left( \Delta l_{s}^{diff} = 0 \right)$. In Equation~\ref{diffusion stress of cell}, $K$ is a coefficient, which depends on Young's modulus ($E$), Poisson's ratio ($\mu$), and the thickness ($l$) of the cathode, separator, and anode (using subscripts `c', `s', and `a', respectively): 
\begin{equation}\label{K value in diffusion stress}
\frac{1}{K}=\sum_{i}^{}\frac{1}{K_{i}},~\text{for } i = c, s,~\text{and}~a, ~
\end{equation}
where
\begin{equation}\frac{1}{K_{i}}=\frac{\left ( 1-2\mu_{i} \right )\left ( 1+\mu_{i}\right )}{E_{i}\left ( 1-\mu_{i} \right )}l_{i}.
\end{equation}
The change in thickness for the anode and cathode ($\Delta l_{c}^{diff}$ and $\Delta l_{a}^{diff}$) can be found from:
\begin{equation}\label{thickness change}
\Delta l_{i}^{diff} = \frac{1}{3}\int \frac{\left ( 1+\mu_i \right )}{1-\mu_{i}}\Omega_i\left ( c_{s} -c_{s}^{o}\right )dx.
\end{equation}
where $\Omega$ is the partial molar volume and the subscripts $i$ is `c' or `a' to represent the cathode or anode, respectively, $c_{s}^{o}$ represents the initial concentration in electrodes, and $c_s$ is the concentration in electrodes at a given time. The partial molar volume $\Omega$ is a property of an electrode that governs the change in volume due to species diffusion, and for LIBs, it depends on the concentration of Li\textsuperscript{+} in the electrodes. For both electrodes (here, the NMC cathode and graphite anode), $\Omega$ is modeled as a function of concentration based on Koerver \textit{et al.} \cite{Koerver2018Chemo-mechanicalBatteries}. Although there is no explicit mention of time ($t$) in Equation \ref{diffusion stress of cell}, $\sigma_{diff}$ is a function of $t$ since the concentration of lithium ($c_s$) varies with time as LIB operates. 
% The diffusion strain can also be found from:
% \begin{equation}\label{diffusion strain}
% \varepsilon_{ij}^{th} = \frac{1}{3}\Delta c\Omega\delta_{ij}.
% \end{equation}
% where  $\delta_{ij}$ is the Dirac delta function and $\Delta c$ is the change in concentration of diffusion species. 
%\Delta l_{s}^{diff}$ is assumed to be zero.

To estimate the \textit{thermal stress} ($\sigma_{th}$), the mechanical equilibrium equation, given by:
\begin{equation}\label{thermal stress}
\triangledown \cdot \sigma_{th} = 0,
\end{equation}
is solved within the cylindrical jellyroll domain. The outer boundaries of the cell (the case) are assumed to be rigid, such that the case prevents any overall deformation. As the temperature ($T$) varies temporally and spatially within the jellyroll (as calculated by the thermal model, see Section~\ref{sec:thermalModel}), this result in thermal strain of:
\begin{equation}\label{Thermal strain of cell}
 \varepsilon_{ij}^{th} = \alpha \Delta T~\delta_{ij}
\end{equation}
where $\alpha$ is the coefficient of thermal expansion and $\Delta T$ is the local temperature change at each location of the domain.

All of the mechanical properties (\textit{i.e.}, $E$, $\mu$ \& $\alpha$) used for the calculation of $\sigma_{diff}$ \& $\sigma_{th}$ are based on an average of the properties within the domain (\textit{i.e.}, cathode, anode, separator and current collectors) that form the cylindrical jellyroll. Table~\ref{Mech_Properties} shows the values of properties and corresponding references from which they are adopted.
% \clearpage
% \subsubsection{\textbf{Coupled Model Implementation}}
% \begin{wrapfigure}[12]{r}{0.5\textwidth}
% \vspace{-10pt}
%     \includegraphics[width=0.5\textwidth]{Battery Model-1/algorithm.png}
%     \caption{ Flowchart detailing the algorithm used to simulated the electrochemical, thermo-fluid, and mechanical response of the battery system.}\label{fig: Flowchart of algorithm}
% \end{wrapfigure}

\begin{figure*}[b!]
     \centering
     \includegraphics[width=\textwidth]{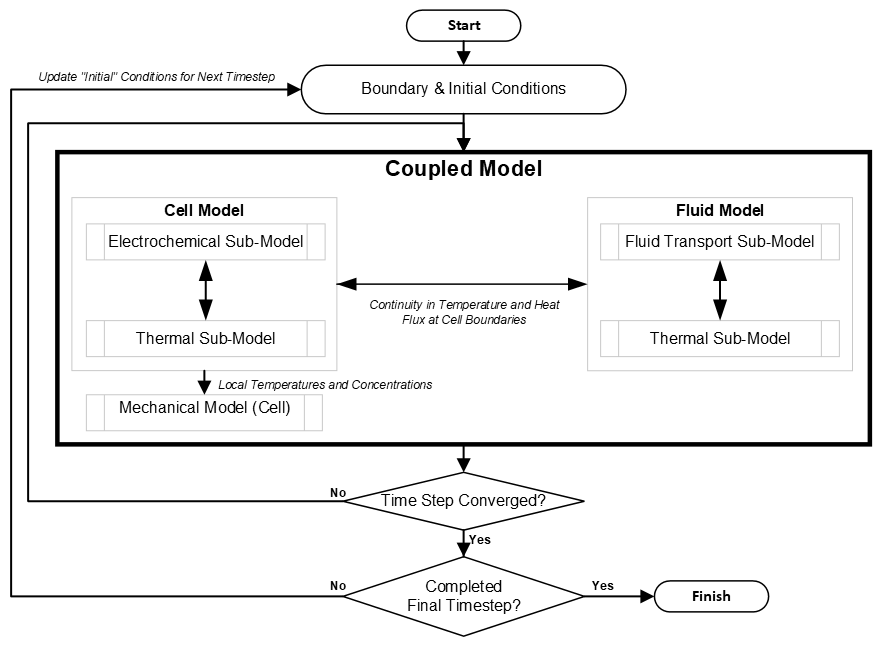}
    \caption{ Flowchart detailing the algorithm for simulating the electrochemical, thermo-fluid, and mechanical response of the battery system. The system solves the coupled model at each time step and then marches forward in time until the cell is fully discharged.}
        \label{fig: Flowchart of algorithm}
\end{figure*}

\subsubsection{\textbf{Coupled Model Implementation}}

As illustrated in the previous sections, the equations that govern electrochemical, thermal, and fluid transport, as well as mechanical stresses, are coupled. Specifically, properties and parameters of P2D electrochemical model depend on temperature (see Equations~\ref{i0}-\ref{electrolyte diffusion ionic}), heat generation is function of outputs from the P2D model intertwined with the temperature (see Equation~\ref{heat generation}), and the temperature within the LIB is governed by the fluid cooling performance (see Equations~\ref{Continuity eqn}-\ref{Energy Conservation eqn}). In other words, the cooling using the dielectric fluid governs the cell domain temperature, which in turn influences the P2D model. Hence, there is significant cross-coupling between different subroutines (\textit{i.e.}, electrochemical, heat transfer, and fluid flow) in the immersion cooling simulations. Due to this two-way coupling, we need to solve the thermal, fluid, and electrochemical sub-models simultaneously while advancing in time, as shown in Figure~\ref{fig: Flowchart of algorithm}. The mechanical model is only one-way coupled with the other models and can be solved after the coupled models converge. In this work, the P2D electrochemical model is solved using the partial differential equation (PDE) interface of COMSOL\textsuperscript{\textregistered}, and the fluid flow and heat transfer models are implemented in the specific COMSOL\textsuperscript{\textregistered} modules. The mechanical model is solved using MATLAB\textsuperscript{\textregistered} after calculation of the other sub-models. The complete algorithm outlined in Figure~\ref{fig: Flowchart of algorithm} is implemented using  LiveLink\textsuperscript{TM} for MATLAB\textsuperscript{\textregistered} that integrates it with COMSOL\textsuperscript{\textregistered}.

In the present study, several discharge rates are analyzed: 1C, 3C \& 5C. For this system, 1C corresponds to a constant current discharge at $I$ = 2.2 A as we are modeling a cell with capacity $CA$ = 2.2 Ah. This matches a recent study from literature \cite{Ji2013Li-IonTemperatures} that is used to validate the model (See Section~\ref{Validation} and Fig.~\ref{fig:Validation}). 
Additionally, we evaluate three different mass flow rates (0.01, 0.005, and 0.0025 kg/s) for the dielectric fluids (mineral oil (`MO') and deionized water (`DIW')). For comparison, we also consider cooling with air at a constant inlet velocity of 0.3 m/s. These liquid flow rates and the air velocity is selected to ensure that the flow is laminar based on the flow rates investigated in the literature \cite{Trimbake2022MineralInvestigation, Jithin2022NumericalFluids}.

\clearpage
\subsection{Model Validation}\label{Validation}
\begin{wrapfigure}[18]{r}{0.5\textwidth}
\vspace{-20pt}
    \includegraphics[width=0.48\textwidth]{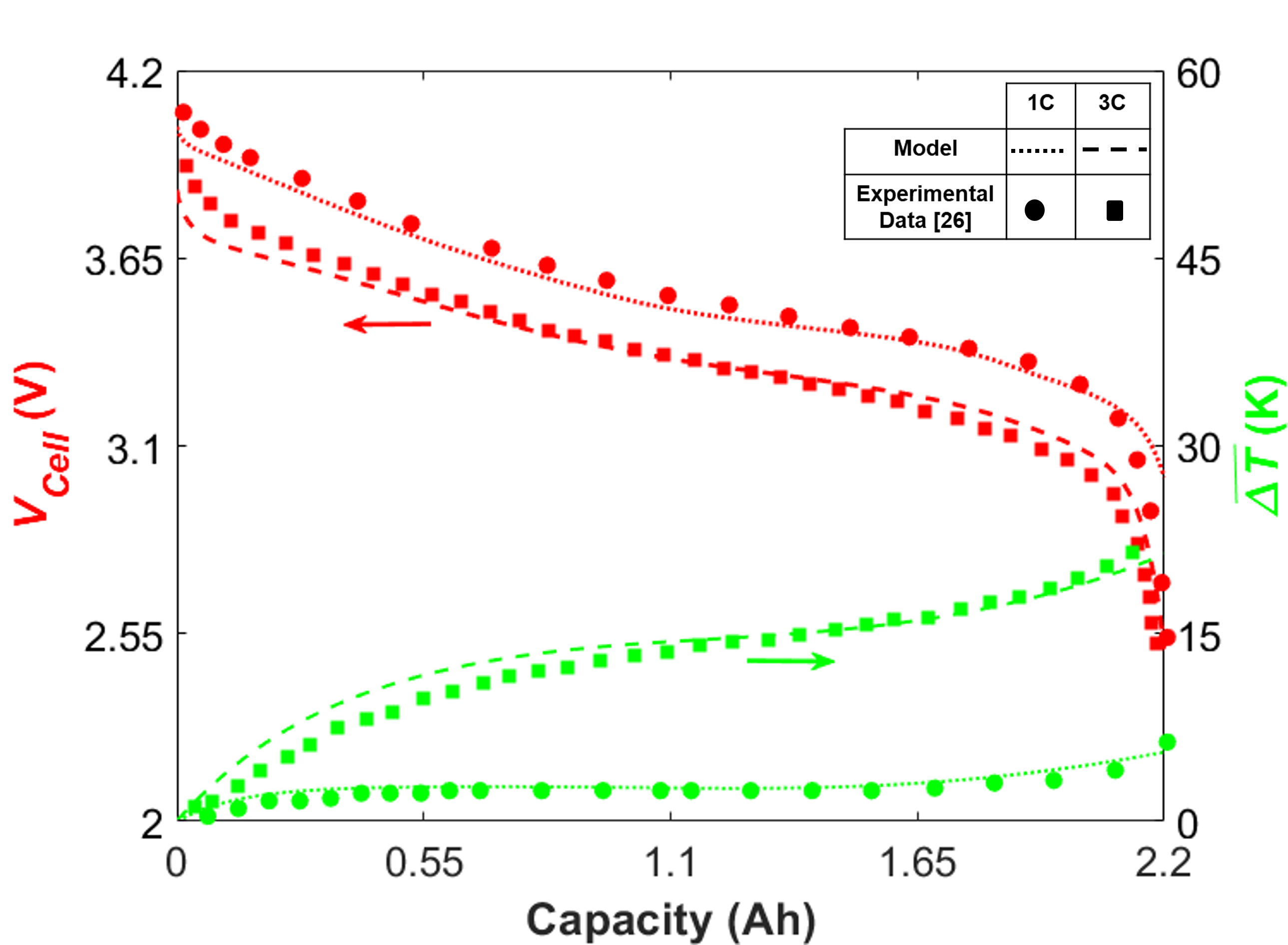}
    \caption{Validation of the fully coupled electrochemical and thermal model with experimental data from Ji \textit{et al.} \cite{Ji2013Li-IonTemperatures} for a case with natural convection to ambient air with a convection coefficient of $h_{nat} \approx$ 28.4~W/(m$^2$K). Comparison of the measured and predicted temperature rise (green) and cell voltage (red) shows that our model accurately predicts the performance of the cell for both 1C and 3C discharge rates. }   \label{fig:Validation}
\end{wrapfigure}

For validating the coupled electrochemical-thermal model, results from the present electrochemical and thermal models are compared with the experimental results from literature for a natural convection cooled battery cell \cite{Ji2013Li-IonTemperatures}. For this validation, the electrochemical model (Section~\ref{P2D_model}) and the thermal model (Section~\ref{sec:thermalModel}) of the cell alone are analyzed \textit{without} the fluid flow and mechanical components. From their experimental data, using a lumped capacitance thermal model, they estimated a natural convection heat transfer coefficient $h_{nat}$ of 28.4 W/m$^2$K, which we then apply at the boundary of our thermal model for the validation of the electrochemical and (conduction) thermal sub-models. Note that the thermophysical properties such as density and specific heat in the lumped model for validation are same as in given in the original paper by Ji \textit{et al.} \cite{Ji2013Li-IonTemperatures}. All properties in the electrochemical model are same are shown in Table~\ref{Tab::P2D model parameters}. 
Figure~\ref{fig:Validation} shows the cell potential ($V_{Cell}$) and the average temperature rise ($\overline{\Delta T}$) for discharge rate of 1C and 3C from our numerically model and the past experimental data. Our numerical model accurately predicts the magnitude of the cell potential $V_{Cell}$ and average temperature rise $\overline{\Delta T}$. 

\clearpage
\subsection{Mesh Independence}
\begin{wrapfigure}[18]{R}{0.5\textwidth}
    \vspace{-15mm}
    \includegraphics[width=0.48\textwidth]{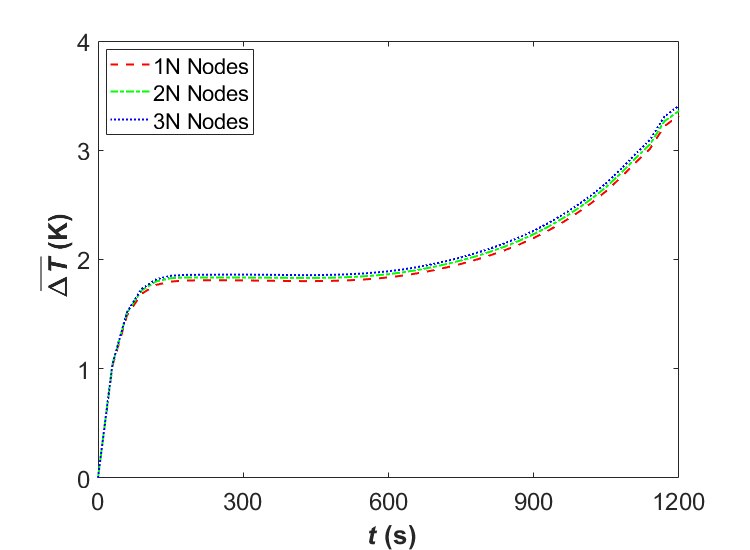}
     \hfill
        \caption{Demonstration of the mesh independence of the full-scale forced immersion cooling model with deionized water for a mass flow rate of 0.01 kg/s at discharge rate of 3C. The temporal evolution of the average temperature rise, $\overline{\Delta T}$, in the cylindrical jellyroll domain  remains approximately constant when doubling or tripling the number of nodes in the computational domain. For the base case, $N \approx$ 400,000 nodes.}  \label{fig:Mesh independence}
\end{wrapfigure}
Numerical simulations can be impacted by number of nodes within the mesh. To ensure that the results are insensitive to the mesh, we evaluate the forced immersion cooling model for deionized water with a mass flow rate of 0.01 kg/s at a discharge rate of 3C. Figure~\ref{fig:Mesh independence} shows the predicted average temperature of the cell, $\overline{\Delta T}$, for three different levels of meshing ($N$, $2N$, and $3N$, where $N \approx$ 400,000 nodes). The total number of nodes includes the full immersion cooling domain (see Figure \ref{fig: Geometry schematic comb} (b)). The results are acceptably similar for all 3 mesh levels. Therefore, the middle level of meshing ($2N \approx 800,000$ nodes) is selected for all further simulations as an optimum choice between computational time and accuracy. 

\vspace{10mm}
\section{Results and Discussion}
Now that we have validated the model against data from literature and evaluated the mesh independence of the model, we use the model to understand the impact of discharge rate and fluid flow parameters on the electrochemical, thermal, and mechanical response of the system. First we analyze the impact of discharge rate in Section~\ref{Effect of discharge rate} and then we consider the varying the mass flow rate in Section~\ref{Effect of mass flow rate}. To further understand the impact of fluid choice on system response, a parametric sensitivity analysis is described in Section~\ref{Sensitivity analysis}. Finally, we define a new metric to accelerate design of immersion cooling systems in Section~\ref{Final Comparison}.
%In the following sections, the results are separated into different subtopics to clearly demonstrate the effect of different operating condition on the FIC performance. Moreover, the results have been displayed from all the subroutines separately in all the subsections. In the end, results from all the configurations have been compared using different metrics. 

\subsection{Effect of Cell Discharge Rate and Fluid Type}\label{Effect of discharge rate}

\begin{figure*}[tb!]
    \centering
    \includegraphics[scale=0.695]{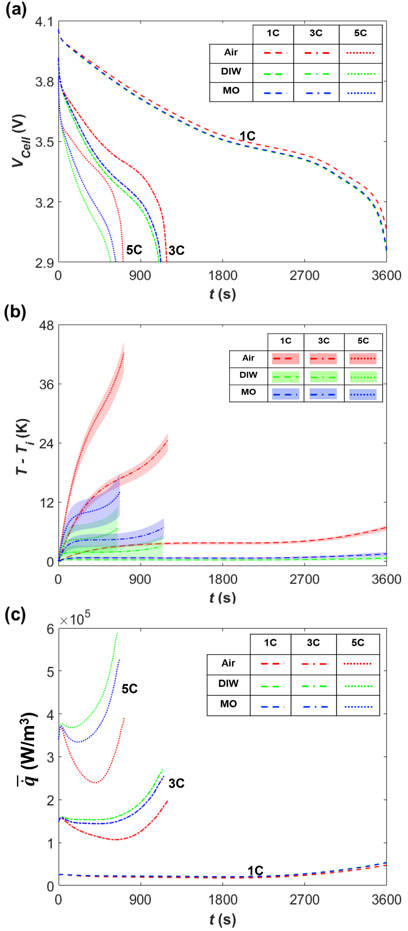}
    \caption{Temporal evolution of (a) the cell potential, $V_{Cell}$; (b) the range of temperature rises, $T-T_i$, in jellyroll region; and (c) the average heat generation rate, $\overline{\Dot{q}}$ for different discharge rates and dielectric fluids. Note that the filled region in panel (b) highlights the range of temperatures between the maximum and minimum temperature within the battery cell and the dashed line is the average temperature rise, $\overline{\Delta T}$. The mass flow rate of both dielectric fluids is 0.01 kg/s and, for comparison, the air inlet velocity is 0.3 m/s.}
    \label{fig: Effect of discharge rate}
\end{figure*}

First, we evaluate effect of discharge rate and fluid type on the cell performance by keeping the mass flow rate of dielectric fluid fixed at 0.01 kg/s and varying the discharge rate (1C, 3C, \& 5C). We compare two immersion cooling liquids: mineral oil (shown in blues throughout the following plots) and deionized water (shown in greens). For comparison to more standard cooling methods, we also evaluate forced convection cooling using air at 0.3 m/s (shown in reds).

\subsubsection{Electrochemical and Thermal Response} As seen in Figure \ref{fig: Effect of discharge rate}, both the type of dielectric fluid and the discharge rate strongly impact both the cell potential and thermal response. This highlights the need for a holistic approach toward understanding immersion cooling, rather than simulating the thermal-fluid aspects in isolation of the electrochemistry. 

Immersion cooling with water (DIW) reduces the average temperature rise ($\overline{\Delta T}$)  significantly compared to the baseline case forced convection cooling by air, with the mineral oil (MO) providing an intermediate response (see Figure~\ref{fig: Effect of discharge rate} (b)). In particular, at the end of the 5C discharge process, the average temperature rise $\overline{\Delta T}$ in the cell is $\sim$40 K for air, $\sim$16 K for MO, and only 8 K for DIW. 

For all fluids, as expected, the magnitude of $\overline{\Delta T}$ increases as the discharge rate increases, since increasingly rapid discharge rates indicates the power within the battery cell is used in a shorter time leading to higher heat generation rates. This trend is clearly demonstrated in Figure~\ref{fig: Effect of discharge rate} (c), which shows the average heat generation ($\overline{\Dot{q}}$) for each case throughout the dsicharge process. Quantitatively, $\overline{\Dot{q}}$ is approximately 5$\times$ and 10$\times$ higher 3C and 5C, respectively, compared to corresponding 1C value. Note that the heat generation rate is not constant in time: as the cell starts to heat up, the heat generation rate first decreases, then it increases as the discharge process nears completion. 

Beyond the average temperature rise in the system, variations in temperature within a cell can impact performance and reliability. Figure \ref{fig: Effect of discharge rate} (b) shows the range of temperature rises above the initial temperature for the different discharge rates and fluids. For all the fluids, the range of temperature variations within the cell increases with increasing discharge rate.  But although improved cooling at the cell boundary (for instance, using liquid water instead of mineral oil) improves the absolute temperatures in the battery cell, the variation within the cell can be large even with high-performance cooling fluids. For example, at 5C, the range of temperatures within the cell is the largest for DIW, followed by MO, and then air. This is opposite of the trend for the average temperature rise (see $\overline{\Delta T}$ in Figure \ref{fig: Effect of discharge rate} (b)). Qualitatively, the Biot Number ($Bi = h L_c / \lambda_{eff}$) provides an estimate of when internal temperature gradients are significant ($Bi>0.1$) based on the effective convection heat transfer coefficient $h$, a characteristic length scale for the cell $L_c$, and the cell thermal conductivity $\lambda$. For a fixed cell with given $L_c$ and $\lambda_{eff}$, the internal temperature gradients will be more likely to be significant when the convection coefficient is large.
Additional discussion on the heat transfer within the cell is included in \hyperref[HeatTransferDetails]{Appendix A}.  

Clearly, the LIB cell performance is influenced by both the type of cooling fluid and the discharge rate. The sensitivity to cooling fluid is exacerbated at higher discharge rates. In other words, at higher discharge rates (\textit{i.e.}, 3C and 5C), all the performance metrics (\textit{i.e.}, $V_{Cell}$, $\overline{\Delta T}$, and $\overline{\Dot{q}}$) differ both in trend and magnitude with the cooling conditions, while at 1C the response, especially that of $V_{Cell}$ and $\overline{\Dot{q}}$, is relatively insensitive to the type of fluid. For example, consider the discharging time ($t_{dis}$) for the LIB cell. The expected trend is that it will take 3600~s, 1200~s, and 720~s to fully discharge the cell at rates of 1C, 3C, and 5C, respectively. At 1C discharge rate, all fluids expend the cell capacity at $\sim$3600~s irrespective of fluid. However, the discharging time at 5C is reduced to $\sim$650~s for DIW-based immersion cooling and $\sim$690~s for MO-based immersion cooling, but matches the expected value of $\sim$720~s when there is only natural convection cooling in air.

 \subsubsection{Mechanical Response}  
 
\begin{figure*}[tb!]
    \centering
    \includegraphics[scale=0.7]{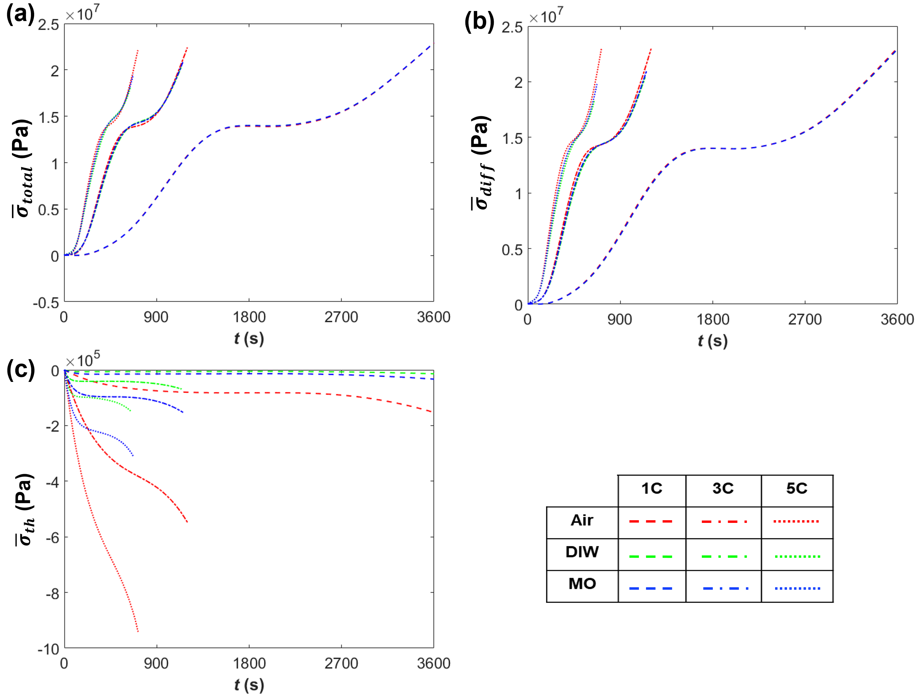}
    \caption{(a) Total stress, $\bar{\sigma}_{total} = \bar{\sigma}_{diff} + \bar{\sigma}_{th}$, (b) Diffusion stress, $\bar{\sigma}_{diff}$,  and (c) Thermal stress, $\bar{\sigma}_{th}$, as a function of time for different dielectric fluid and discharge rate. These values are the average stress within the cell domain during the discharge process. The total stress $\bar{\sigma}_{total}$ follows the diffusion stress $\bar{\sigma}_{diff}$ closely since the magnitude of $\bar{\sigma}_{diff}$ is two order larger than the $\bar{\sigma}_{th}$. Note that the legend applies to all panels. Also, the stress value represent change in the stress from t = 0 s in the cell domain over the course of the discharge process.}
    \label{fig:  stress in discharge rate}
\end{figure*}

 After predicting the temperature and electrochemical response, the mechanical response of the cell is determined for the same set of cooling fluids and cell discharge rates. Figure \ref{fig: stress in discharge rate} shows the value of average diffusion stress ($\bar{\sigma}_{diff}$), average thermal stress ($\bar{\sigma}_{th}$) and total stress ($\bar{\sigma}_{total} $), which is summation of $\bar{\sigma}_{diff}$ and $\bar{\sigma}_{th}$. 
 Also, the diffusion stress is two orders of magnitude larger than the thermal stress throughout the discharging process. As a result $\bar{\sigma}_{total}$ trend (see Figure \ref{fig: stress in discharge rate}) is similar to the $\bar{\sigma}_{diff}$ plot. In other words, the net stress, $\bar{\sigma}_{total}$ within the cell domain is tensile and its magnitude is same as $\bar{\sigma}_{diff}$.  
 
 The trend of the temporal evolution of the diffusion stress is similar for all immersion cooling conditions (see Figure~\ref{fig: stress in discharge rate} (a-b)). Specifically, the plateau region in the middle of the discharge process, as opposed to the monotonic behavior, is related to the value of $\Omega$ for each electrode (see Ref. \cite{Koerver2018Chemo-mechanicalBatteries}). Clearly, temporal stretch of the plateau region in $\bar{\sigma}_{diff}$ decrease with increasing discharge rate for all the fluids (\textit{i.e.}, DIW, MO, \& Air), which could be related to the high speed of diffusion that eventually leads to fast variation in $c_s$. Also, note that the diffusion stress is a tensile stress, which indicates that NCM electrode expands less than the graphite electrode shrinks during the discharge process. The diffusion stress is sensitive to the immersion cooling fluids only during a short period in the middle of the discharge process. Even at 5C, the diffusion stress is generally independent of the cooling fluid.
 
 In contrast to the diffusion stress, the thermal stress (Figure~\ref{fig: stress in discharge rate} (c)) is significantly influenced by the discharge rates as well as the type of fluids (see Figure \ref{fig: stress in discharge rate} (c)), since its magnitude is directly proportional to the $\overline{\Delta T}$ (see Figure \ref{fig: Effect of discharge rate} (b)). Thus, the thermal stresses are the highest for 5C discharge with air as the cooling fluid and lowest for DIW at 1C discharge. However, the thermal stresses are generally two orders of magnitude smaller than the diffusion stresses and thus there is minimal impact on the overall stress response. Note that the stress calculated in the analysis represent the stress generated (or change) from t = 0 s and the absolute value depends on the initial value of stress at the beginning of the discharge process.

\subsection{Mass Flow Rate Dependence}\label{Effect of mass flow rate}

\begin{figure*}
     \centering
     \begin{subfigure}[b]{0.49\textwidth}
         \centering
         \includegraphics[width=\textwidth]{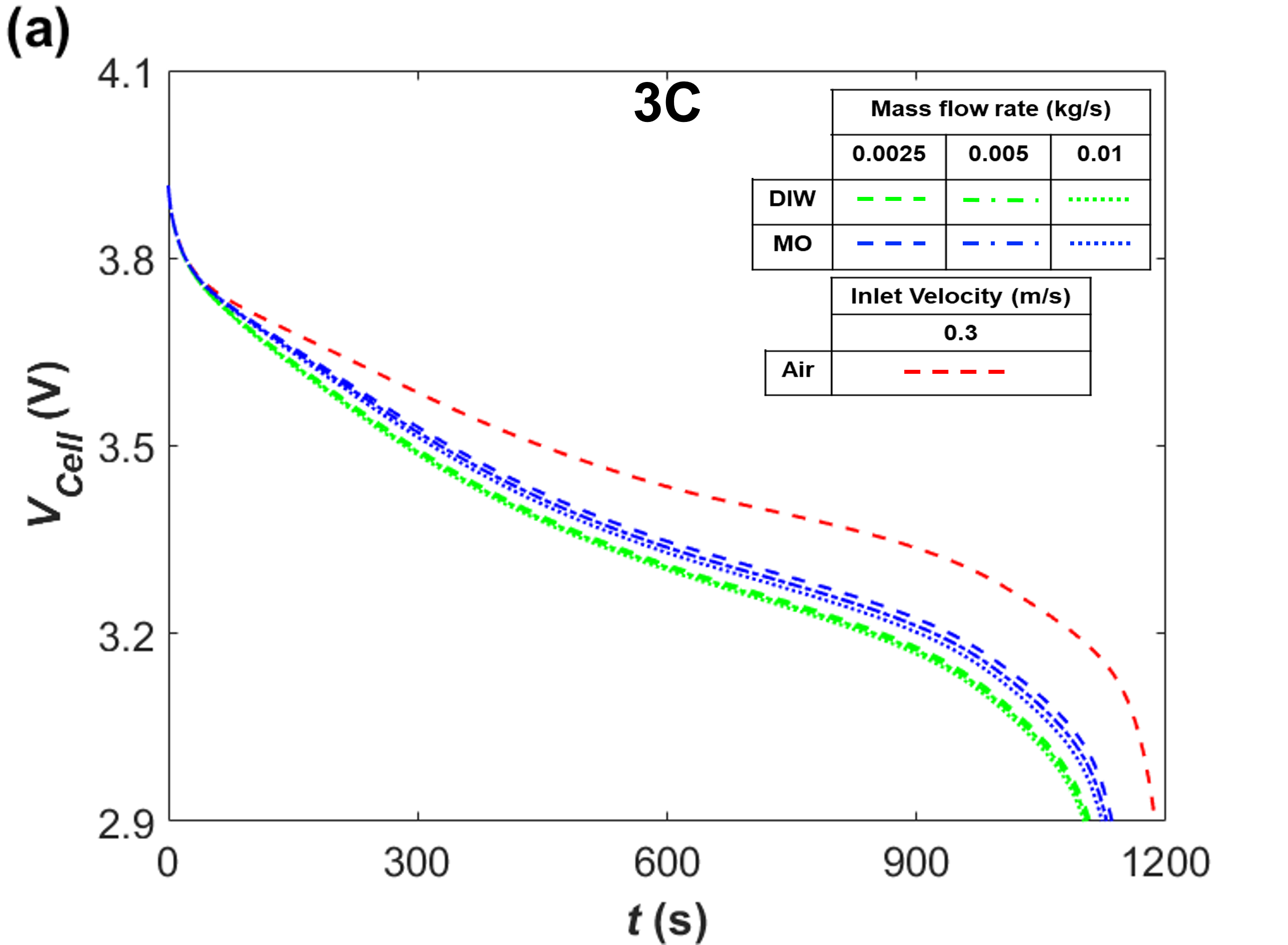}
         %\caption{}
         \label{Fig. 7(a)}
     \end{subfigure}
     \hfill
     \begin{subfigure}[b]{0.49\textwidth}
         \centering
         \includegraphics[width=\textwidth]{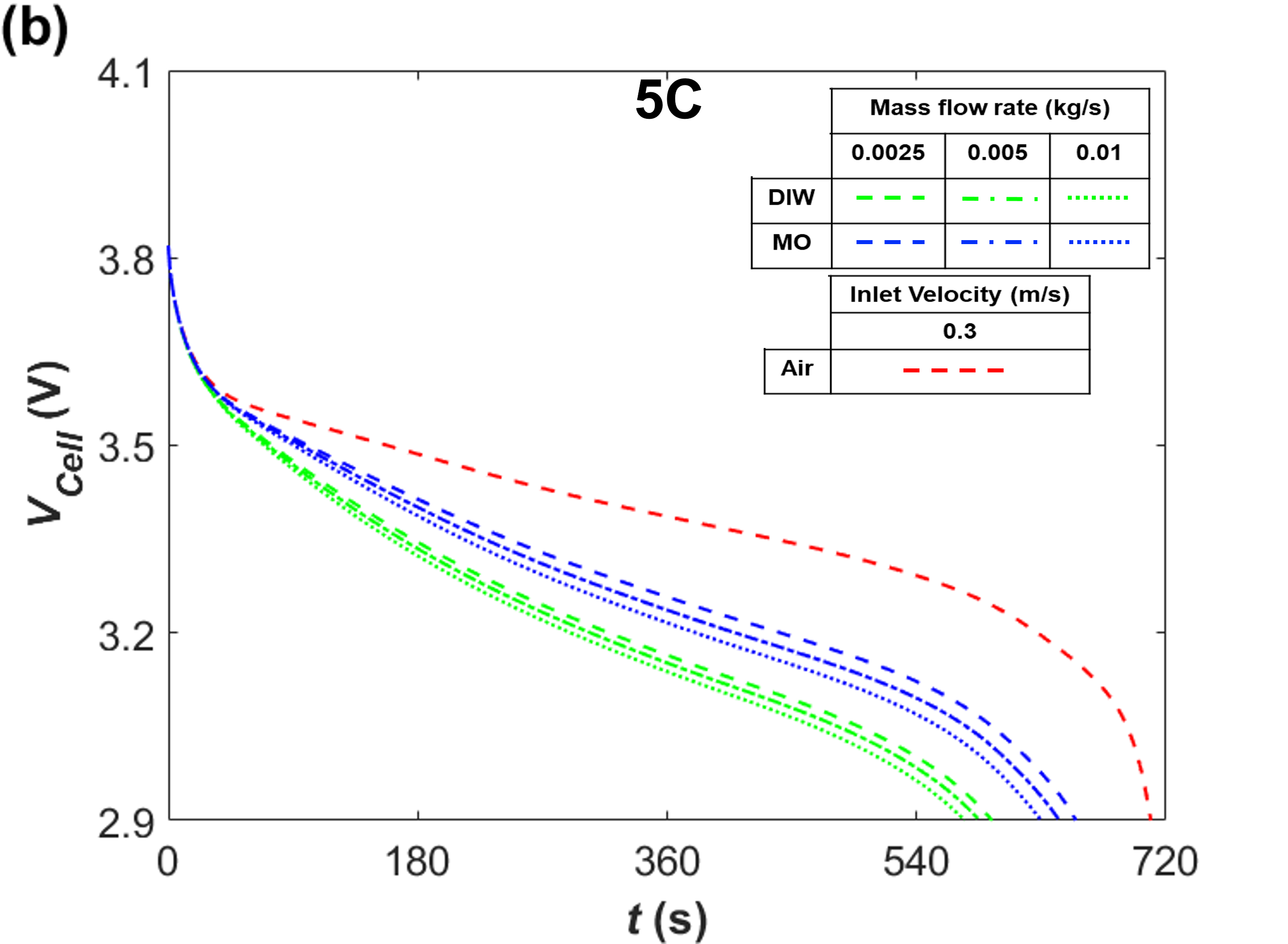}
         %\caption{}
         \label{Fig. 7(b)}
     \end{subfigure}
     \hfill
      \begin{subfigure}[b]{0.49\textwidth}
         \centering
         \includegraphics[width=\textwidth]{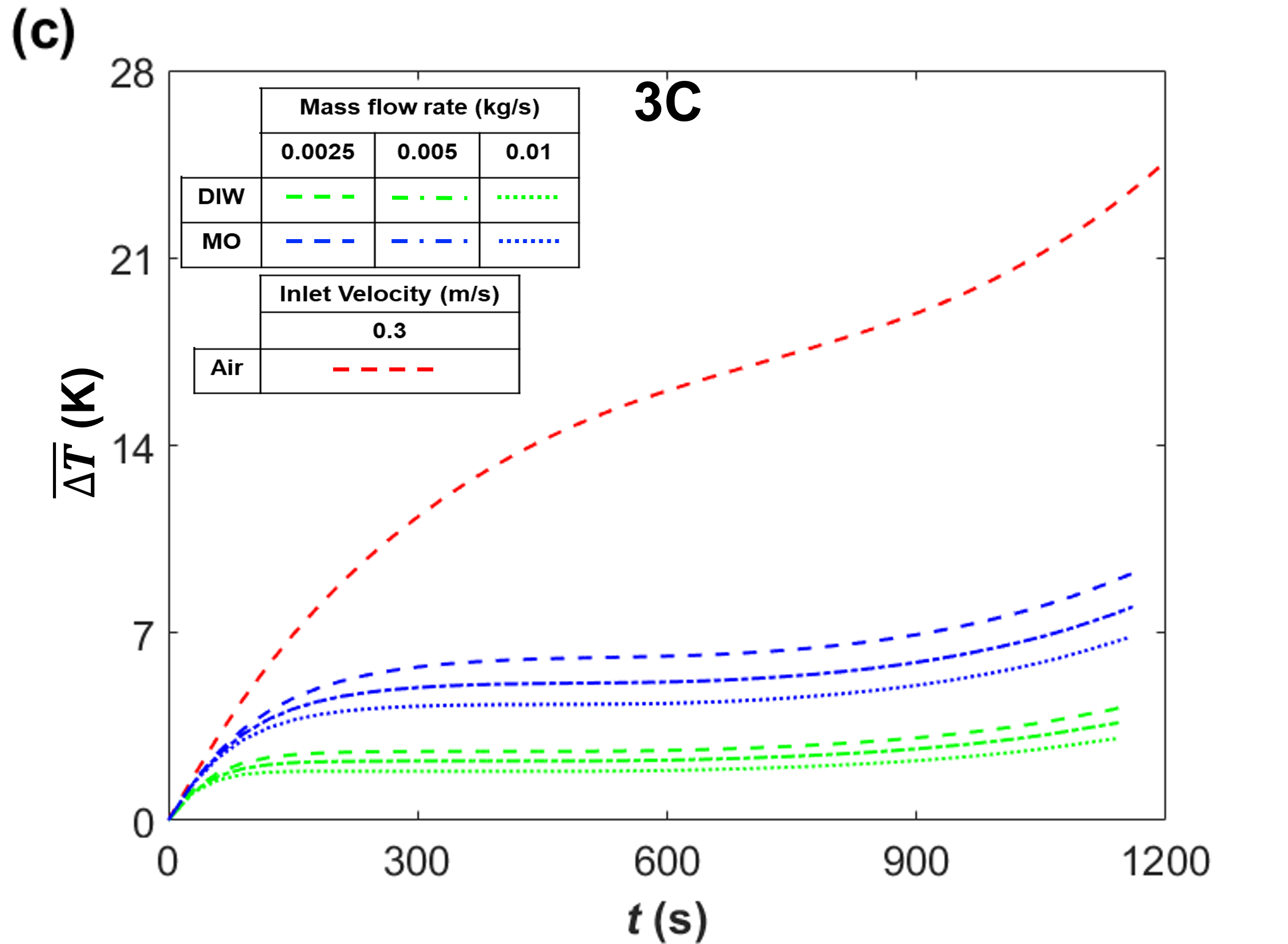}
         %\caption{}
         \label{Fig. 7(c)}
     \end{subfigure}
      \hfill
     %\hfill
     % \begin{subfigure}[b]{0.5\textwidth}
     %     \centering
     %     \includegraphics[width=\textwidth]{Battery Model-1/tmaxmin_1_3C.png}
     %     %\caption{}
     %     \label{Fig. 7(d)}
     % \end{subfigure}
     % \hfill
%         \caption{ Temporal evolution of (a) Cell voltage, $V_{Cell}$, (b) Average temperature rise, $\overline{\Delta T}$ in the cylindrical jellyroll domain, and (c) Average heat generation rate, $\overline{\Dot{q}}$ in the cylindrical jellyroll domain for different mass flow rate of dielectric fluids at a \textbf{discharge rate of 3C} as mentioned in legends. %Note that the filled region in sub-figure (d) highlights the difference between the maximum and minimum temperature, $T_{max}-T_{min}$ in the cell domain and the dashed line is the average temperature. 
%         The air inlet velocity is fixed at 0.3 m/s in the comparison case.}
%         \label{fig: Effect of mass flow rate-3C}
% \end{figure*}
% \begin{figure*}
     \centering
     \begin{subfigure}[b]{0.49\textwidth}
         \centering
         \includegraphics[width=\textwidth]{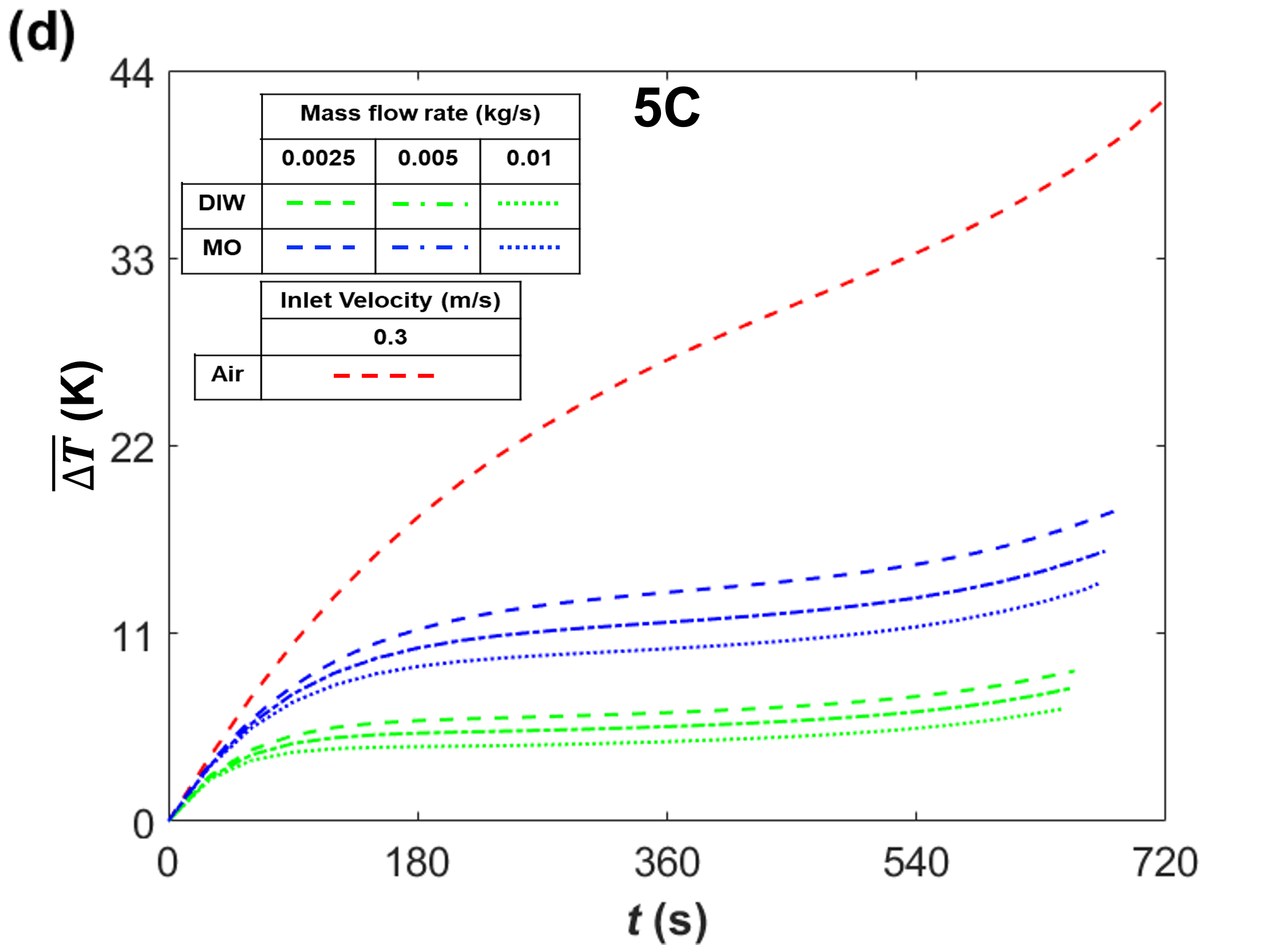}
         %\caption{}
         \label{Fig. 7(d)}
     \end{subfigure}
     \hfill
     \begin{subfigure}[b]{0.49\textwidth}
         \centering
         \includegraphics[width=\textwidth]{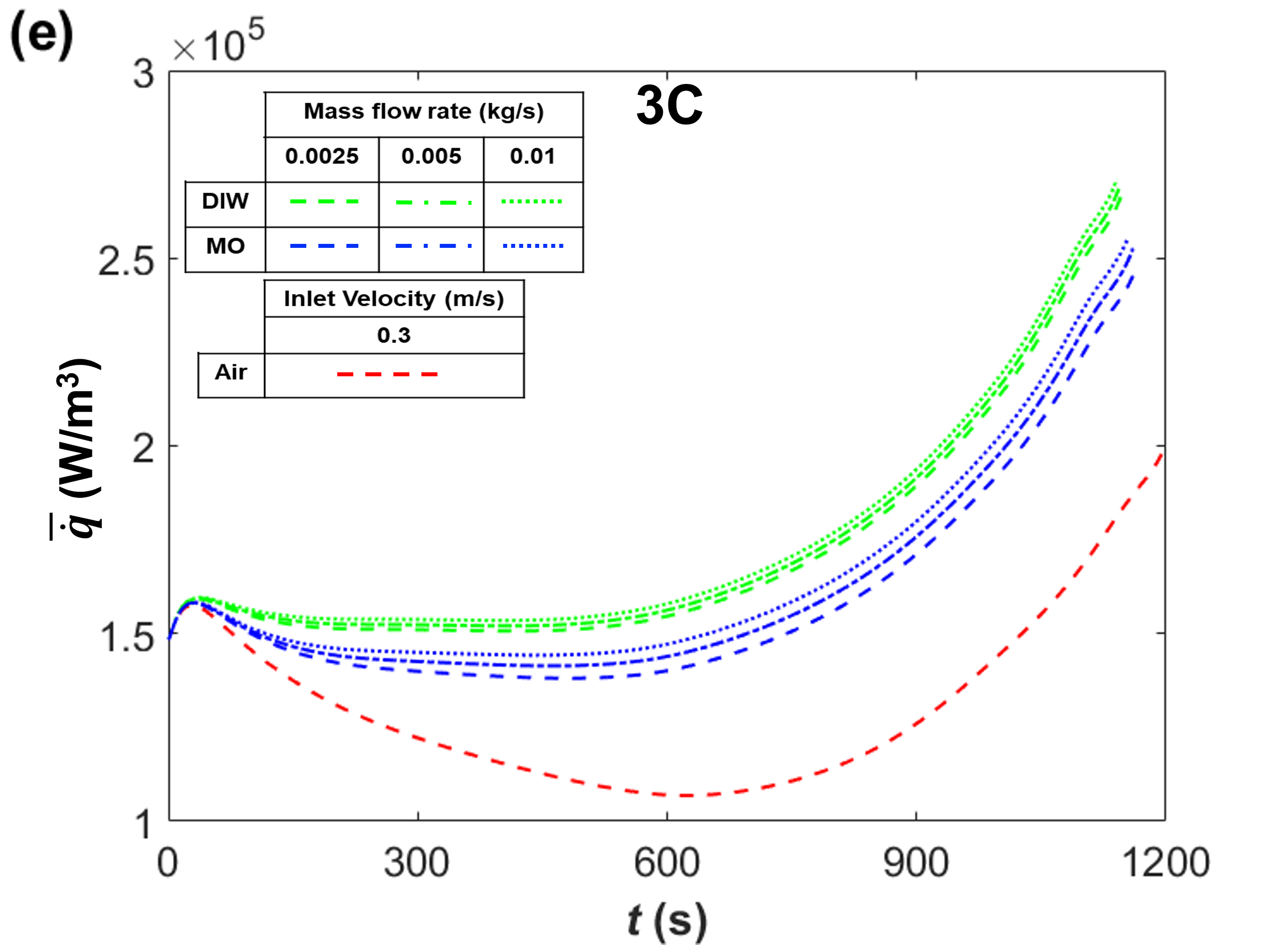}
         %\caption{}
         \label{Fig. 7(e)}
     \end{subfigure}
     \hfill
      \begin{subfigure}[b]{0.49\textwidth}
         \centering
         \includegraphics[width=\textwidth]{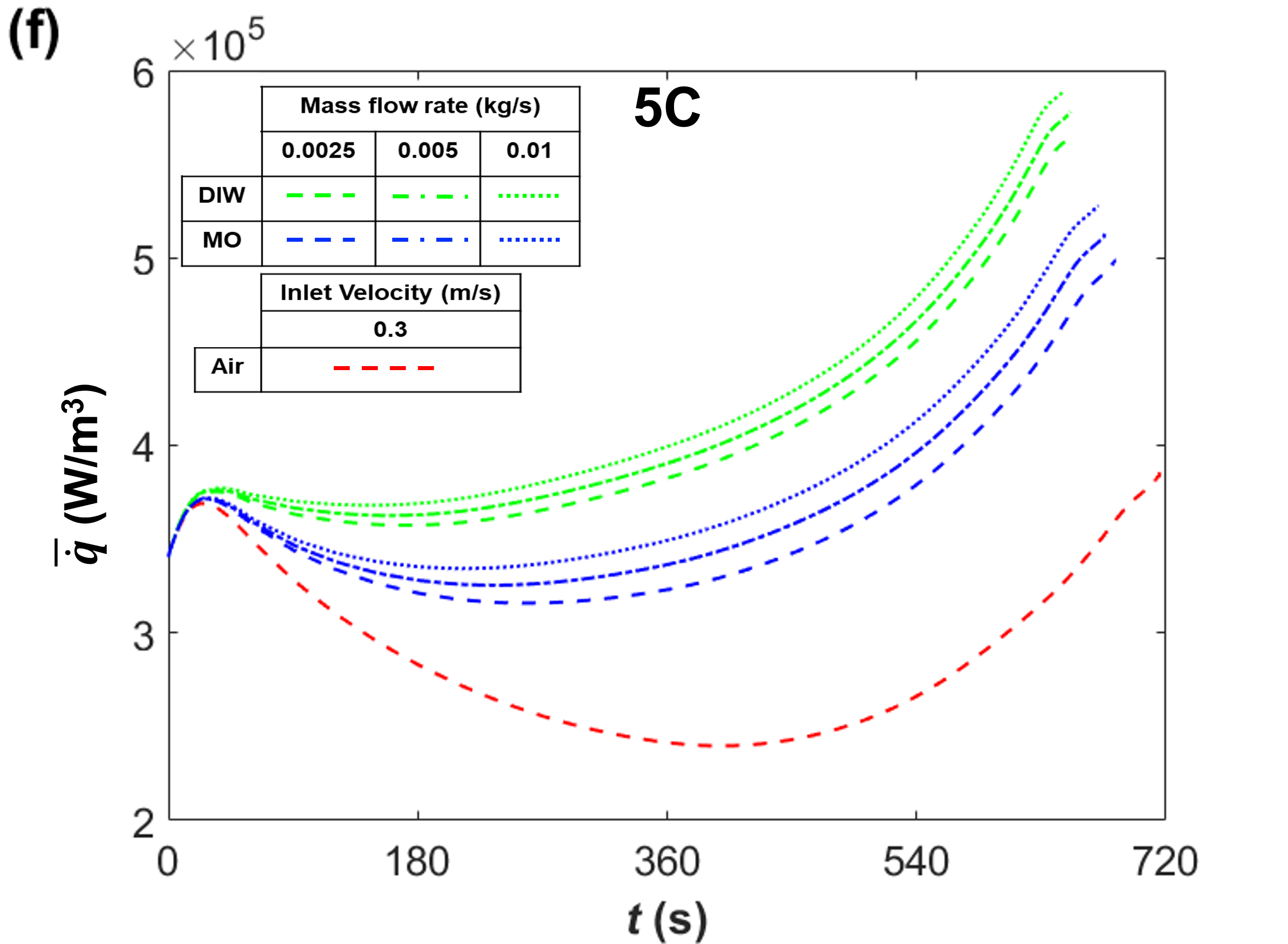}
         %\caption{}
         \label{Fig. 7(f)}
     \end{subfigure}
      \hfill
     % \hfill
     % \begin{subfigure}[b]{0.5\textwidth}
     %     \centering
     %     \includegraphics[width=\textwidth]{Battery Model-1/Tmaxmin_1_5C.png}
     %     %\caption{}
     %     \label{Fig. 6(d)}
     % \end{subfigure}
     \hfill
        \caption{Temporal evolution of (a-b) the cell voltage, $V_{Cell}$, (c-d) average temperature rise, $\overline{\Delta T}$, in the cylindrical jellyroll domain, and (e-f) average heat generation rate, $\overline{\Dot{q}}$, in the cylindrical jellyroll domain at (a,c,e) 3C and (b,d,f) 5C discharge rates for different mass flow rate of the dielectric fluids. %Note that the filled region in sub-figure (d) highlights the difference between the maximum and minimum temperature, $T_{max}-T_{min}$ in the cell domain and the dashed line is the average temperature. 
        The air inlet velocity is fixed at 0.3 m/s in the comparison case.}
        \label{fig: Effect of mass flow rate}
\end{figure*}
The mass flow rate of the dielectric fluids is one of the key operating parameters for an immersion cooling system and impacts the electrochemical performance of the system due to the coupling to the thermal response. To elucidate the effect of mass flow rate for our system, we consider three different mass flow rates (\textit{i.e.}, 0.01, 0.005, and 0.0025 kg/s) for two cell discharge rates (\textit{i.e.}, 3C and 5C). Results for liquid immersion cooling are compared to air cooling at an inlet velocity of 0.3 m/s.

\subsubsection{Electrochemical and Thermal Performance} Several key conclusions can be drawn from the electrochemical and thermal results for immersion cooling with different fluids and mass flow rate combinations (see Figures~\ref{fig: Effect of mass flow rate} for results at 3C and 5C.). 

First, 
we can see that as the mass flow rate decreases, the magnitude of $\overline{\Delta T}$ increases for both dielectric fluids (\textit{i.e.}, DIW \& MO). This follows the expected trend since convective cooling is directly proportional to mass flow rate. However, the thermal performance of MO-based immersion cooling configuration is more sensitive to the variation in mass flow rate compared to DIW because of the inherently higher cooling capacity (as discussed in Section \ref{Effect of discharge rate}). 

Second, the evolution of the cell potential, $V_{Cell}$, is also influenced by the mass flow rate as shown in panel (a) \& (b) of figure \ref{fig: Effect of mass flow rate}. However, similar to average temperature rise, $\overline{\Delta T}$, the effect of mass flow rate on the cell potential is more apparent as the mass flow rate decreases and the discharge rate increases. 

Third, the range of temperatures within the cell domain does not follow a straightforward monotonic trend with increasing mass flow rate at either discharge rate.  \hyperref[HeatTransferDetails]{Appendix A} provides a more detailed discussion on the temperature variation within the cell. 

Lastly, despite the average temperature rise being the lowest for DIW-based configuration with the highest mass flow rate (0.01 kg/s), it has the highest rate of heat generation, $\overline{\Dot{q}}$, as shown in Figure \ref{fig: Effect of mass flow rate}(e) and (f). In fact, the average heat generation rate is the lowest for the air-cooled configurations $\overline{\Dot{q}}$, which has the highest temperature rise. Overall, there is a clear trend (here and in Section \ref{Effect of discharge rate}) that lower average temperature rise leads to larger heat generation rates. Initially, this seems somewhat counter-intuitive (if one considers that more heat generation ties to higher temperatures \textit{assuming} a fixed convection coefficient for heat losses and uniform and constant material properties). However, here, the temperature and heat generation rate are closely coupled. Specifically, the high heat generation at lower temperatures results from the temperature dependence of the electrical resistance of the cell. Due to the electrochemical kinetics, the cell resistance decreases with increasing temperature in this temperature regime. More insight into this thermal behavior is provided in \hyperref[HeatTransferDetails]{Appendix A}. This explanation is further backed by the observation that the DIW-based cooling suffers from a higher capacity loss ($CA_{loss} = CA - It_{dis}$, when compared to the other two fluids (MO and Air), especially at high discharge rates (3C and 5C). In other words, LIB cells cooled with DIW discharge quicker than other cooling configurations (see also Section \ref{Effect of discharge rate}). The capacity loss follows a clear monotonous inverse trend with respect to average temperature, $\overline{\Delta T}$, for all the configurations at a given discharge rate (increasing $\overline{\Delta T}$ reduces the apparent capacity loss). This effect is also important for battery cells operating at low ambient temperatures \cite{Ji2013Li-IonTemperatures,Hao2018EfficientAlloy,Wang2016AVehicles, Lyu2020InvestigationStorage}. 
 
To summarize, the highly cross-coupled and nonlinear nature of the immersion-cooled battery system impacts heat transfer and electrochemical performance. Therefore, as the battery starts discharging, the temperature rise is lower for high-performance cooling fluids in immersion cooling configurations. This results in higher electrical resistance of the cell, which in turn leads to higher $\overline{\Dot{q}}$ compared to less effective cooling cases. However, due to the better heat dissipation for the higher-performance fluid, the temperature rise remains lower. %But since the fluid can carry more heat results i  and as has been of in all the configurations the important is a clear nonlinear nature of the FIC configurations. 
%This further bolsters the main motivation of the present study made in Introduction (see Section \ref{Introduction}) that all the physical phenomena within the FIC system (\textit{i.e.}, electrochemical, thermal, fluid flow) should be studied in conjunction for a realistic perspective. 

\subsubsection{Pressure Drop and Pumping Power} One other critical parameter of \textit{forced} immersion cooled systems is pressure drop ($\Delta p$)  required to drive the flow at the required mass flow rate. This parameter governs the pumping power ($P_{pump}$) required to operate the cooling system. The pressure drop depends strongly on the type of fluid (and its underlying properties such as viscosity) and the flow rate. Note that the fluid property is assumed to be independent of temperature in the present study since the domain only consist of single cell and therefore pressure drop is insensitive to cell discharge rate for these fluids. However, the temperature dependency may becomes important for immersion cooled BTMS with large number of cells. Table~\ref{Tab::Pressure drop} shows the pressure drop ($\Delta p$) and pumping power ($P_{pump}$) for DIW and MO at three different flow rates in the immersion cooling domain. The highest $\Delta p$ and $P_{pump}$ is for the mineral oil at the highest flow rate, while the lowest $\Delta p$ and $P_{pump}$ water at the slowest flow rate. This is due to the fluid viscosity: $\mu_f$ for MO is approximately 50$\times$ greater than that of DIW. In all cases, the pumping power is less than 150 $\mu$W. This corresponds to less than %100~mJ 
150 $\mu$Wh of energy for pumping during the discharging process of 1C. For comparison, one 18650 battery cell stores approximately $\sim 10$Wh\cite{Quinn2018EnergyCells, Abbott2023ExperimentalMethods}. Thus, for a single cell, the pumping power is not a major concern.

\begin{table}[h!]
    \caption{Pressure drop and pumping power for immersion cooling with DIW and MO at different flow rates}
    \begin{center}
    \begin{tabular} { |p{3cm}|p{1.5cm}|p{3cm}|p{1.5cm}|p{3cm}|}
         \hline
         \multirow{2}{*}{\textbf{Mass Flow Rate} (kg/s)} & \multicolumn{2}{|c|}{\textbf{DIW}} & \multicolumn{2}{|c|}{\textbf{MO}} \\
         \cline{2-5}
         & $\Delta p$ (Pa) & $P_{pump}$ ($\mu$W) & $\Delta p$ (Pa) & $P_{pump}$ ($\mu$W) \\
          \hline
         0.01  & 0.332  & 3.32 & 11.315 & 122.4\\
          \hline
        0.005 & 0.125  & 0.62 & 5.658 & 30.61\\
           \hline
         0.0025 & 0.055  & 0.13 & 2.829 & 7.65\\
           \hline           
    \end{tabular} \label{Tab::Pressure drop}
    \end{center}
\end{table}

\subsection{Sensitivity Analysis: Cooling Fluid Properties}\label{Sensitivity analysis}

\begin{figure*}
     \centering
     \begin{subfigure}[b]{0.49\textwidth}
         \centering
         \includegraphics[width=\textwidth]{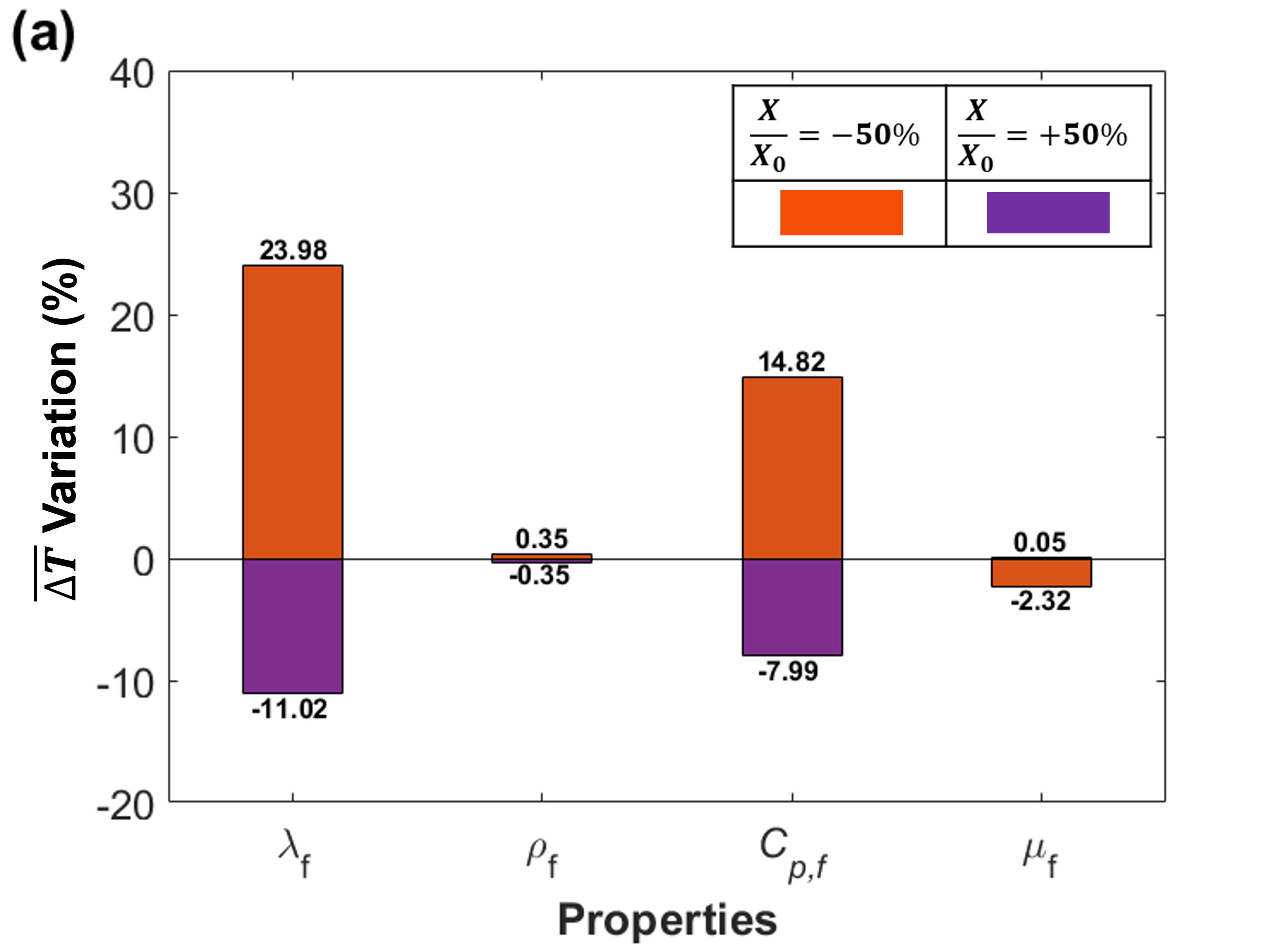}
         %\caption{}
         \label{Fig. 8(a)}
     \end{subfigure}
     \hfill
     \begin{subfigure}[b]{0.49\textwidth}
         \centering
         \includegraphics[width=\textwidth]{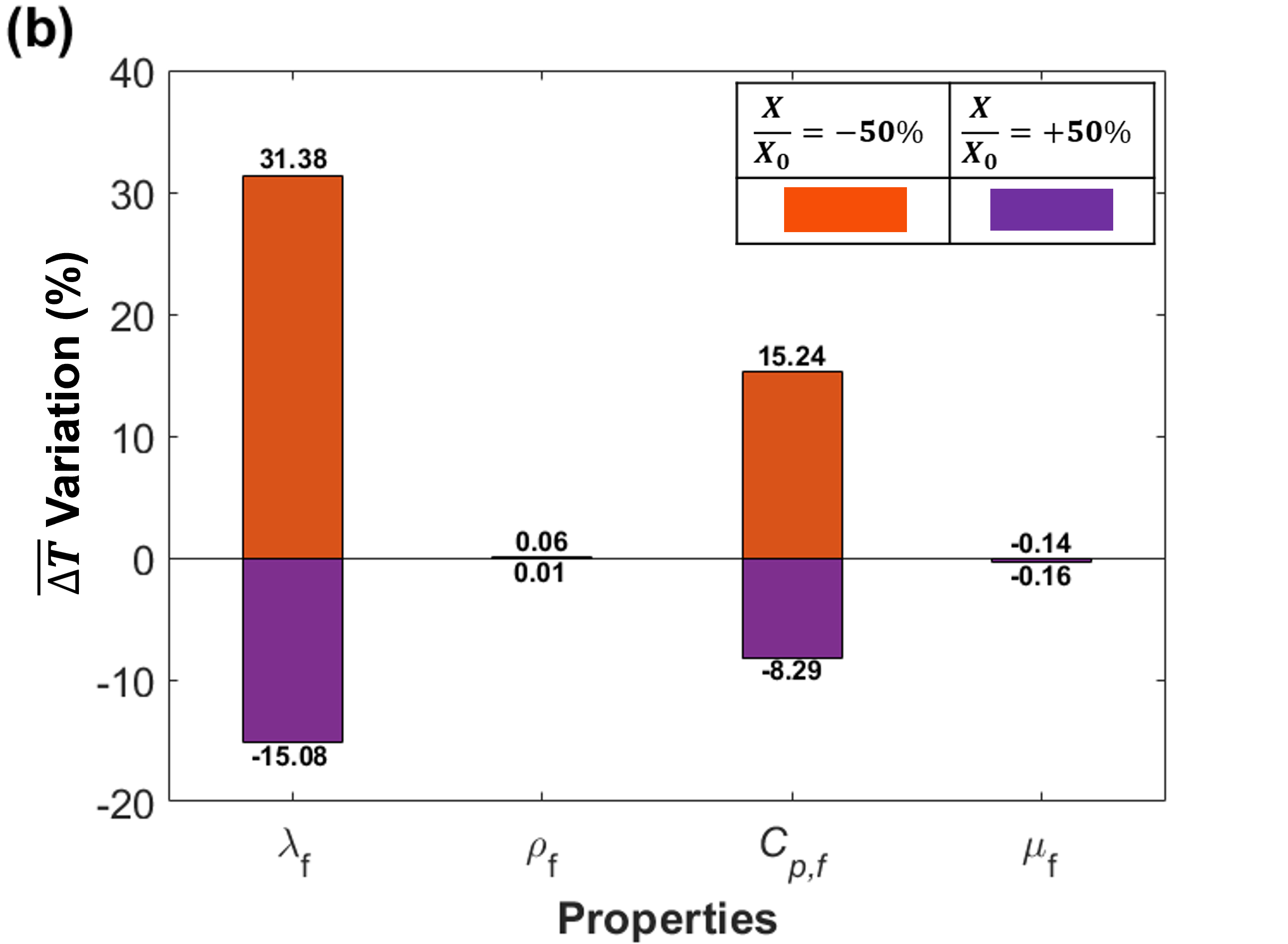}
         %\caption{}
         \label{Fig. 8(b)}
     \end{subfigure}
     \hfill
      \begin{subfigure}[b]{0.49\textwidth}
         \centering
         \includegraphics[width=\textwidth]{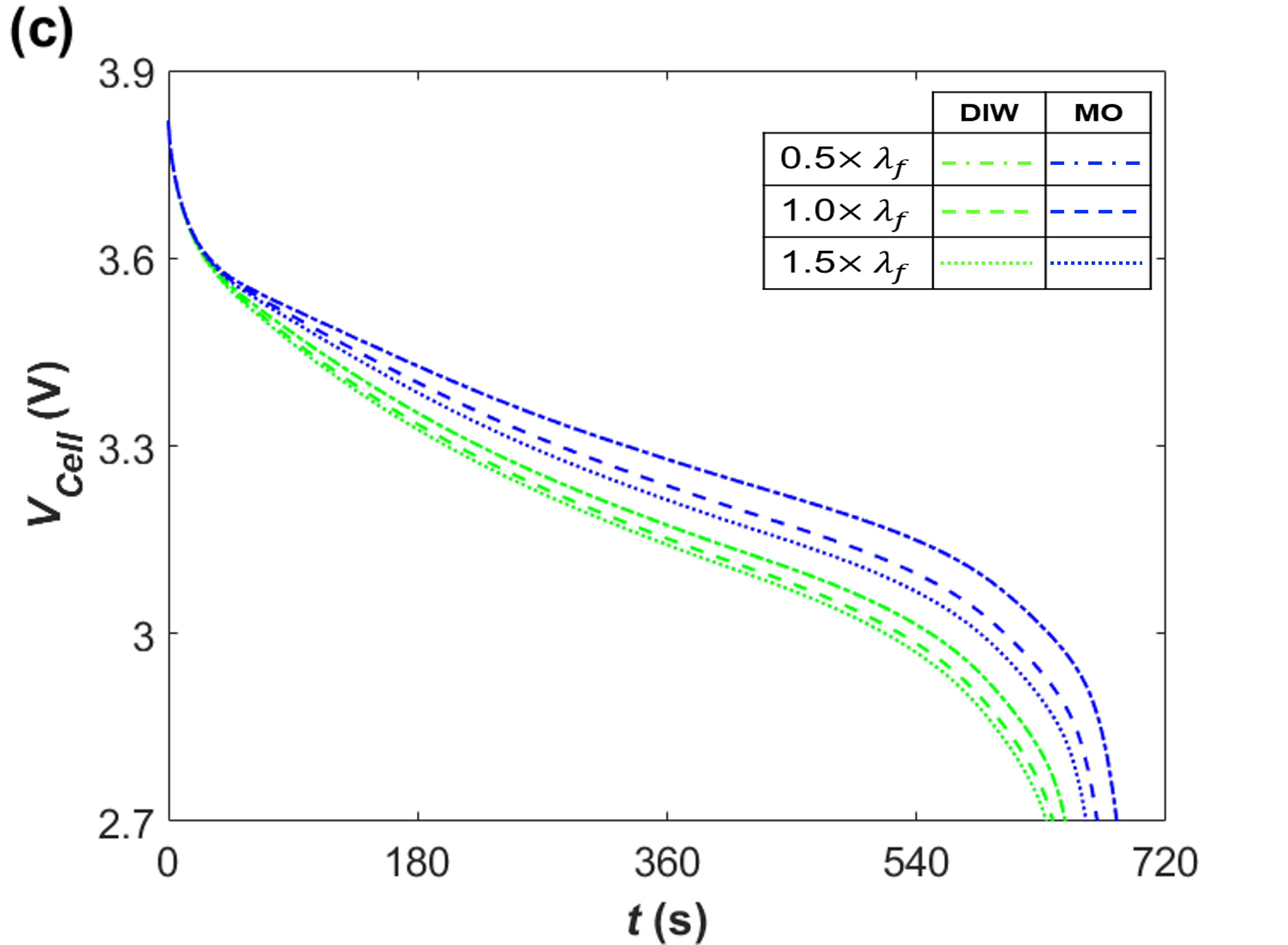}
         %\caption{}
         \label{Fig. 8(c)}
     \end{subfigure}
     \hfill
     \begin{subfigure}[b]{0.5\textwidth}
         \centering
         \includegraphics[width=\textwidth]{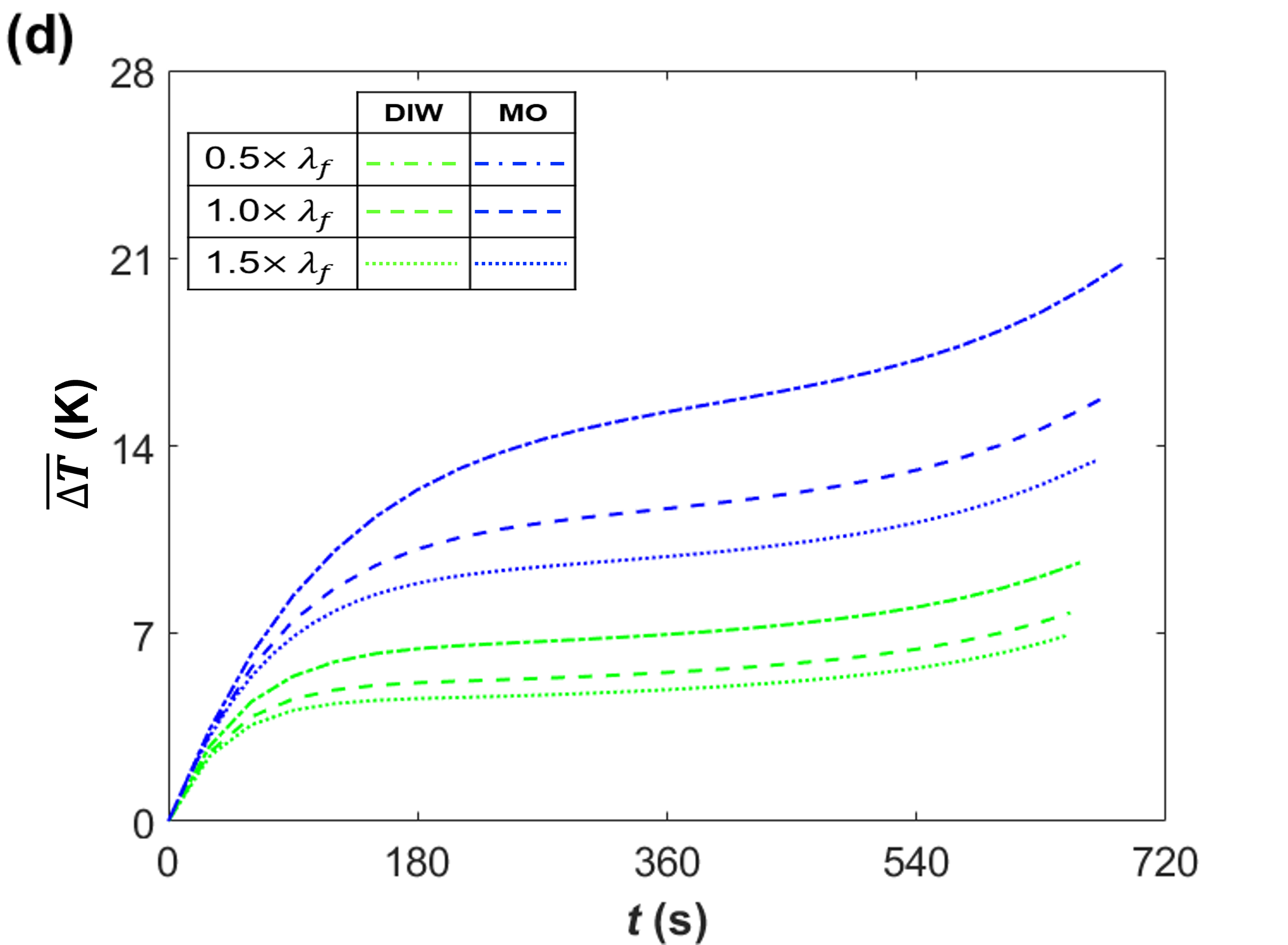}
         %\caption{}
         \label{Fig. 8(d)}
     \end{subfigure}
     \hfill
        \caption{\textit{Sensitivity analysis to thermophysical properties of the cooling fluid}: Percentage change in the average temperature rise with respect to the nominal fluid performance for (a) DIW and (b) MO. The red bars are for the 50\% decrease and the purple bars are for the 50\% increase in the property. 
        Temporal evolution of (c) cell voltage, $V_{Cell}$ and (d) average temperature rise, $\overline{\Delta T}$ in the cylindrical jellyroll domain from the sensitivity analysis when varying thermal conductivity of the fluid, $\lambda_f$. %The green and blue dashed lines are the results with the nominal properties of the DIW and MO. 
        The green and blue dashed lines for 1.0$\times\lambda$ are the results with the nominal properties of the DIW and MO. Note that all other set of results fall within the two extremes, that correspond to the $\lambda$ results (which shows highest variation: highest $\overline{\Delta T}$ for 0.5$\times\lambda$ and lowest $\overline{\Delta T}$ for 1.5$\times\lambda$).  Note that sensitivity analysis is performed for 5C discharge rate and mass flow rate of 0.005 kg/s.}
        \label{fig: Sensitivity analysis}
\end{figure*}

In the previous sections, clearly, the DIW provides better thermal control than the MO. However, given that multiple properties are different between the DIW and MO, the influence of the underlying thermophysical properties of the dielectric fluids on the system performance remains unclear. To address this, we perform a sensitivity analysis by independently varying the individual properties of fluids that contribute to the cooling capacity and response of the system. Specifically, all the fluid properties [such as density ($\rho_f$), thermal conductivity ($\lambda_f$), specific heat ($C_{p,f}$), and viscosity ($\mu_f$)] are varied one at a time considering a 50\% increase and 50\% decrease with respect to original properties, while other properties remain at the nominal values. So for $\lambda_f$ sensitivity two set of properties are simulated: (0.5$~\lambda_f$, $\rho_f$, $C_{p,f}$, and $\mu_f$) and (1.5$~\lambda_f$, $\rho_f$, $C_{p,f}$, and $\mu_f$), where the terms with the subscript `f' represent the original properties of fluid. This results in 8 combinations of FIC configuration for both DIW and MO to compare with the baseline fluid. For this analysis, we consider a discharge rate of 5C with a mass flow rate of 0.005 kg/s for both dielectric fluids. This case provides the most significant variation in performance for the slight change in fluid properties. 

Figure \ref{fig: Sensitivity analysis} shows the results of this sensitivity analysis for both fluids. Generally, the thermal conductivity of the fluid, $\lambda_f$, is the most critical property of a dielectric fluid followed by $C_{p,f}$ in an immersion cooling configuration from the thermal aspect (at a given mass flow rate). For example, for DIW a 50\% decrease in $\lambda_f$ results in $\sim$24\% increase in $\overline{\Delta T}$ and a 50\% increase in $\lambda_f$ leads to $\sim$11\% decline $\overline{\Delta T}$, whereas 50\% decrease in $C_{p,f}$ results in $\sim$15\% increase in $\overline{\Delta T}$ and a 50\% increase in $C_{p,f}$ leads to $\sim$8\% decline $\overline{\Delta T}$. However, similar variations in the other two properties ($\rho_f$ \& $\mu_f$) do not result in a significant change in the thermal response of the system for both DIW- and MO-based immersion-cooled configurations. This trend is impacted by fixing the mass flow rate (at 0.005 kg/s for the sensitivity analysis) as opposed to fixing the fluid velocity or fixing the allowed pressure drop. This keeps the thermal response nearly unaltered (with respect to the original DIW and MO systems) with variations in $\rho_f$ and $\mu_f$. For example, reducing the $\rho_f$ by 50\% will approximately double the average velocity to maintain the same mass flow rate, which does not significantly alter the cooling performance of the fluid. Nevertheless, it is worth pointing out that other flow parameters such as pressure drop ($\Delta p$) will be significantly affected by varying $\rho_f$ and $\mu_f$. 
 
Furthermore, Figure \ref{fig: Sensitivity analysis} (c) and (d) show the change in the $V_{Cell}$ and $\overline{\Delta T}$ response from the sensitivity analysis, where the bounds correspond to (0.5~$\lambda_f$, $\rho_f$, $C_{p,f}$, $\mu_f$) and (1.5~$\lambda_f$, $\rho_f$, $C_{p,f}$, $\mu_f$) for both the dielectric fluids. Note that the lower thermal conductivity (0.5~$\lambda_f$) corresponds to a higher temperature and lower capacity loss. This again highlights the inverse relation between the temperature rise and capacity loss (see Section~\ref{Effect of mass flow rate}). Moreover, the 50\% change in thermal conductivity for the MO-based fluid leads to more significant variation in the response compared to the DIW. To summarize, the superior thermal performance of DIW can be attribute to its higher $\lambda_f$ and $C_{p,f}$, which are approximately 4$\times$ and 2$\times$ higher than that of MO.

\subsection{Metric for Comparing Immersion Cooling Configurations}\label{Final Comparison}
As mentioned in Section~\ref{Introduction}, analyzing immersion cooling requires detailed numerical simulations. There is no universal metric that can be used to compare different combinations of mass flow rate and type of fluid with different discharging rates. Such an analytical metric could be very helpful in the design and operation/control of immersion cooling systems. Often empirical heat transfer correlations are developed for particular flow configurations, and correlations exist for a range of fluids and operating conditions (for example, forced vs. natural convection, internal vs. external flow, and different geometries). Such correlations can become the foundation for a metric to estimate cooling configuration performance without detailed numerical simulations. To begin, we consdier a heat transfer correlation for a cylinder in crossflow \cite{Zukauskas1972HeatCrossflow, Morgan1975TheCylinders}, which has been updated to account for the cylinder in duct flow \cite{Pederson1977HeatFlow}. This correlation predicts for Nusselt number ($Nu = h_{cor}D/\lambda_f$) then predicts the heat transfer coefficients:
\begin{equation}\label{heat transfer correlation}
    h_{cor} = \frac{\lambda_f}{D}~\left ( 0.655Re^{0.471}Pr^{\left ( 1/3 \right )}\left ( 1+\sqrt{D/W} \right ) \right )
\end{equation}
where $Re=\frac{\rho_f D \overline{v}}{\mu_f }$ is the Reynolds number based on the cell diameter ($D$) and average inlet velocity ($\overline{v}$), $Pr= \mu_f C_{p,f}/\lambda_f$ is the Prandtl number, and $W$ is the width of the duct ($W = 2~\Delta x_{perp}~+~D$ in the present analysis). Note that ($1+\sqrt{D/W}$) accounts for the effect of duct and cylinder geometry.

\begin{figure*}
     \centering
     \begin{subfigure}[b]{0.49\textwidth}
         \centering
         \includegraphics[width=\textwidth]{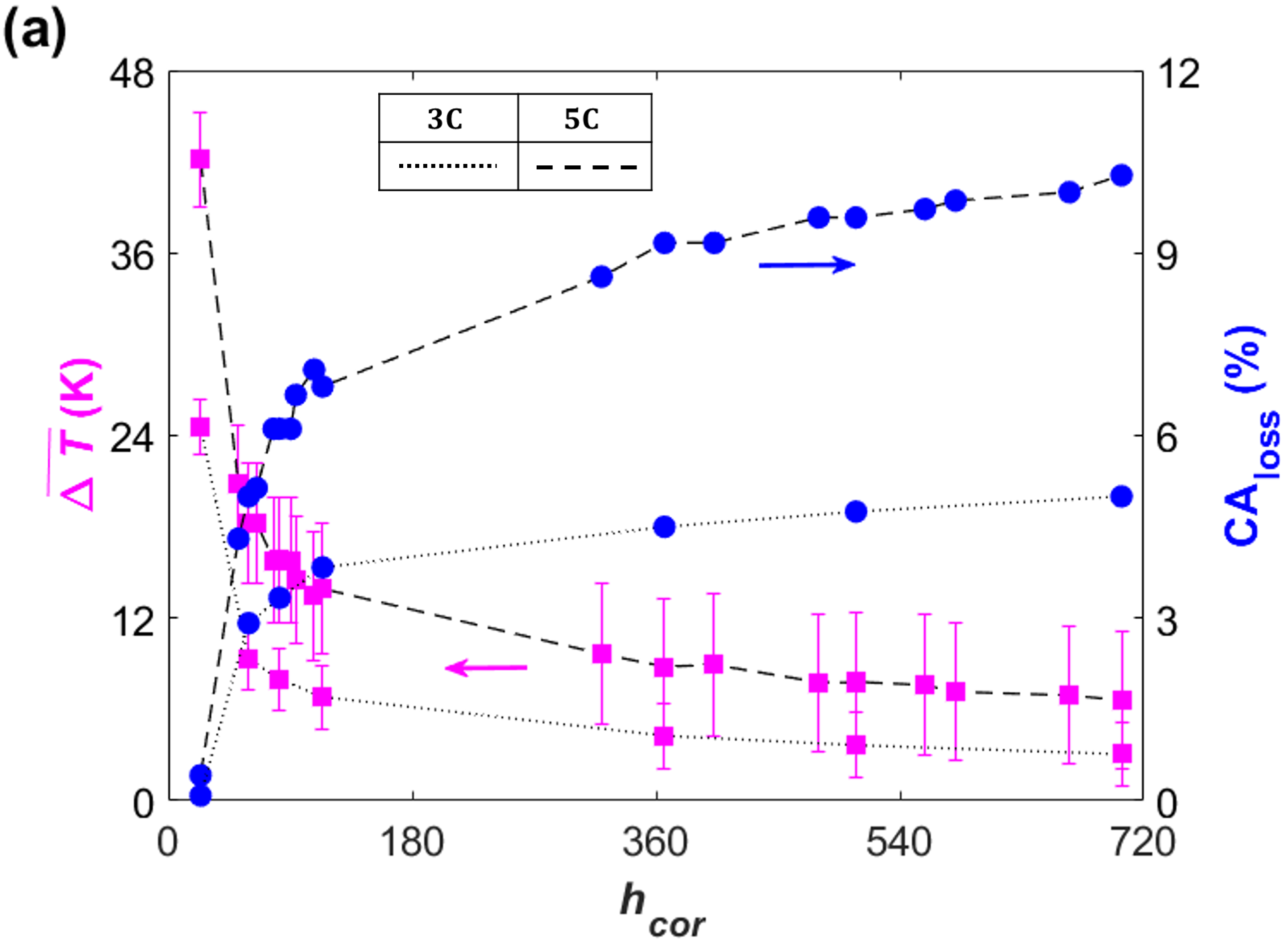}
         %\caption{}
         \label{Fig. 9(a)}
     \end{subfigure}
     \hfill
     \begin{subfigure}[b]{0.49\textwidth}
         \centering
         \includegraphics[width=\textwidth]{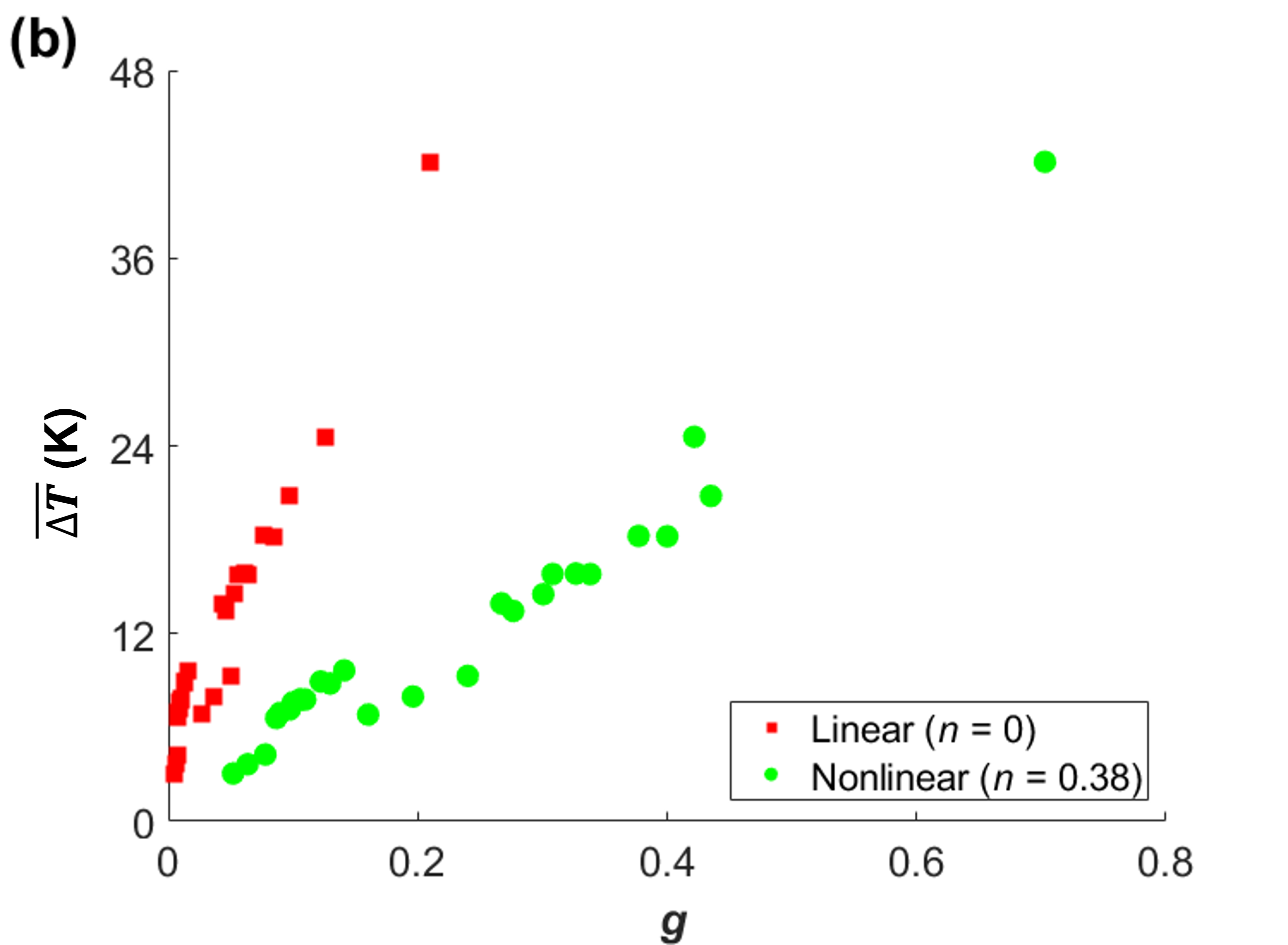}
         %\caption{}
         \label{Fig. 9(b)}
     \end{subfigure}
     \hfill
        \caption{(a) Comparison of the average temperature rise ($\overline{\Delta T}$) and the capacity loss ($CA_{loss}$) for all the forced immersion cooling configurations [including various mass flow rates, types of fluid, and discharge rates] as a function of the convection heat transfer rate, $h_{cor}$ calculated based on an existing correlation (Eq.~\ref{heat transfer correlation}). There is a clear monotonic trend for both parameters with respect to this convection correlation. Note that the ``error bars'' for $\overline{\Delta T}$ represent the range of temperatures within the cell. (b) Average temperature rise as a function of the newly proposed metric, $g$ (see Eq. \ref{Updated heat transfer correlation}) for all the immersion cooling conditions at both discharge rates (3C and 5C) using a linear ($n = 0$, shown in red squares) and nonlinear ($n = 0.38$, shown in green circles) formulation.}
        \label{fig: Final Comparison}
\end{figure*}

Figure~\ref{fig: Final Comparison} (a) shows the average temperature rise and capacity loss for all the cases spanning 3 mass flow rates (0.0025, 0.005, and 0.01~kg/s) and 19 combinations of different fluid properties (DIW, MO, Air, and 16 hypothetical fluids from the sensitivity analysis). This corresponds to  7 data points for 3C and 24 for 5C discharge. For each discharge rate, there is a monotonic decrease in the temperature rise and a monotonic increase in the capacity loss as the convection coefficient increases. This demonstrates that the convection correlation can be a metric that can be used for analytical estimates on which cooling fluids flow rates will perform well at a given discharge rate. %It also sums up the cross-coupled nature of electrochemical and heat transfer behaviour as has been observed in all the previous sections. In other words, designing a FIC configurations that provides a best thermal control (\textit{i.e.}, lowest $\overline{\Delta T}$) may be the worst in terms of electrochemical performance (\textit{i.e.}, highest $Ca_{loss}$) and vice versa. For example, air has the higher $\overline{\Delta T}$ and lower $Ca_{loss}$ whereas DIW has lower $\overline{\Delta T}$ and higher $Ca_{loss}$. So, an ideal FIC system should be designed and operated such that it achieves an optimum balance between the thermal and electrochemical performance.

 In order to extend the versatility, the variation with cell discharge rate should be included in the proposed. One possible approach is to include an estimate of the heat generation rate as it is a direct function of the discharge rate and is linked to temperature response. However, the heat generation rate is not a quantity that is known accurately before computational analysis, so we can make some estimates for a simplified metric. From a lumped capacitance heat transfer analysis, we can link the temperature rise and the heat generation rate as $\overline{\Delta T}\sim (\overline{\Dot{q}} V)/(h_{cor} A_s)$, where $V$ is the cell volume and $A_s$ is the surface area in contact with the cooling fluid.  If we assume that the heat generation rate scales linearly with discharge rate ($\overline{\Dot{q}}$ is independent of cooling performance), we can estimate $\overline{\Dot{q}} \sim P~R$, where $R$ is the discharging rate (e.g., 5 for 5C) and $P$ is a function of LIB cell properties, which is an NCM cell in the present study.
 If we include some non-linearity (\textit{i.e.}, $\overline{\Dot{q}}$ is dependent on cooling), then we can estimate $\overline{\Dot{q}} \sim P~R~h_{cor}^n$, with a power law dependence on the convection coefficient.  
 Mathematically, these can be consolidated into a single form: 
 \begin{equation}\label{Updated heat transfer correlation}
    \overline{\Delta T} \sim P~R~h_{cor}^{n-1} = P~g,
\end{equation}
 where we define a new metric $g = R~h_{cor}^{n-1}$ to estimate system performance. For the linear case, $n = 0$, and for the nonlinear case $n \approx 0.38$, based on the best fit of the data in Figure~\ref{fig: Final Comparison}(a). Figure \ref{fig: Final Comparison}(b) shows the temperature rise for all immersion cooling configurations as a function of this proposed metric assuming $P$ is constant for a given LIB cell. Although there is some scatter in the data, a clear monotonic increase in temperature rise with respect to $g$ is apparent. Thus, $g = R~h_{cor}^{n-1}$ is an effective metric to compare thermal performance of different combinations of fluids, flow rates, and cell discharge rates. % and one of advantage with nonlinear formulation is that there is high sensitivity of $\overline{\Delta T}$ with respect to $g$.

\section{Conclusion}
\noindent In the present study, we analyze the electrochemical, thermal, and mechanical response of an 18650 battery cell with different immersion cooling fluids and flow rates using a fully-coupled modeling approach. Specifically, the detailed pseudo-2D electrochemical model is solved in conjunction with a 3D thermo-fluid model, and results are integrated with a mechanical model to predict stresses in the cell. 

From the analysis of different fluids, flow rates, and discharge rates, the average cell temperature rise follows a trend opposite to that of the capacity loss and average heat generation rate due to the coupled effects of temperature and electrochemical response. Due to the improved electrochemical kinetics with increasing temperature in the temperature range of these simulations, cells operating at lower temperatures due to improved cooling dissipate more heat than those operating at higher temperatures. In turn, the discharging rate is impacted by the temperature rise and higher temperatures retain more battery capacity for these moderate temperature rises. Note that at higher temperatures, catastrophic effects can occur when the heat generation rate increases with increasing temperature leading to thermal runaway.  
%Interestingly, unlike the $\overline{\Delta T}$ temperature distribution within the LIB domain does not show a general increasing or decreasing trend across different FIC configurations. This highlights the multiphysics and nonlinear aspect of FIC, where heat transfer and electrochemical phenomena influence each other. However, the cross-coupled nature of FIC is prominent at higher discharge rates like 3C \& 5C. 
Considering the coupled thermal and electrochemical effects, a holistic understanding of the forced immersion cooling system is necessary for the efficient design and operation of such a system. 

The sensitivity analysis, considering varying the dielectric fluid properties, reveals that forced immersion cooled system thermal performance is most sensitive to thermal conductivity and specific heat of the fluid (for a fixed mass flow rate and cell discharge rate). Therefore, fluids with high $\lambda_f$ and $C_{p,f}$ will provide superior temperature control for similar operating conditions. This is evident in the results that illustrate that DIW leads to lower average temperature rises compared to MO. 

Finally, starting from a correlation for the convection heat transfer coefficient $h_{cor}$, we propose and demonstrate the applicability of a metric $g = R~h_{cor}^{n-1}$ that can rank the performance of different cooling fluids and flow rates across different cell discharge rates $R$. There is a clear monotonic decrease in the temperature rise and increase in capacity loss as $h_{cor}$ rises at a given discharge rate. The proposed metric extends this analysis to include the impact of discharge rate on cell response. Now, with a single analytical equation, potential fluids for immersion cooling can be ranked and the impact of changing parameters can be analyzed.
Combining the analytical metric to screen potential fluids with the multi-physics numerical models can make the battery thermal management system design more efficient. As mentioned in the introduction (see Section \ref{Introduction}), although the present study focuses on discharging process, all the finding (like temperature control, sensitivity analysis and use of metric) is valid for immersion cooling based BTMS of LIB batteries independent of charging/discharging operation.

%Although there is an approximate monotonic increase in the average temperature rise with this metric, there is still room to improve the correlation. One potential extension of the present work is to further improve the formulation of $g$ or to develop an analytical method to compare different immersion cooling configurations without detailed multi-scale/multi-physics simulations. Another future direction would be to reduce the computational cost associated with the fully-coupled models. By integrating analytical models for certain ``physics'',  to provide an optimum trade-off between accuracy and computational cost. Alternatively, data-driven statistical and machine-learning approaches can be used to analyze such complex systems efficiently. Beyond the parameters varied in his study, there is a need to evaluate the impact of cell chemistry and geometry, as well as building up from a single cell to a battery pack or module. 

\section*{Acknowledgment}
This material is based upon work supported by the National Science Foundation under Grant No. 2143043. Any opinions, findings, and conclusions or recommendations expressed in this material are those of the authors and do not necessarily reflect the views of the National Science Foundation.

\section*{Appendix A: Details of Heat transfer within the LIB cell}\label{HeatTransferDetails}

\begin{figure*}[tbh!]
     \centering
     \begin{subfigure}[b]{0.49\textwidth}
         \centering
         \includegraphics[width=\textwidth]{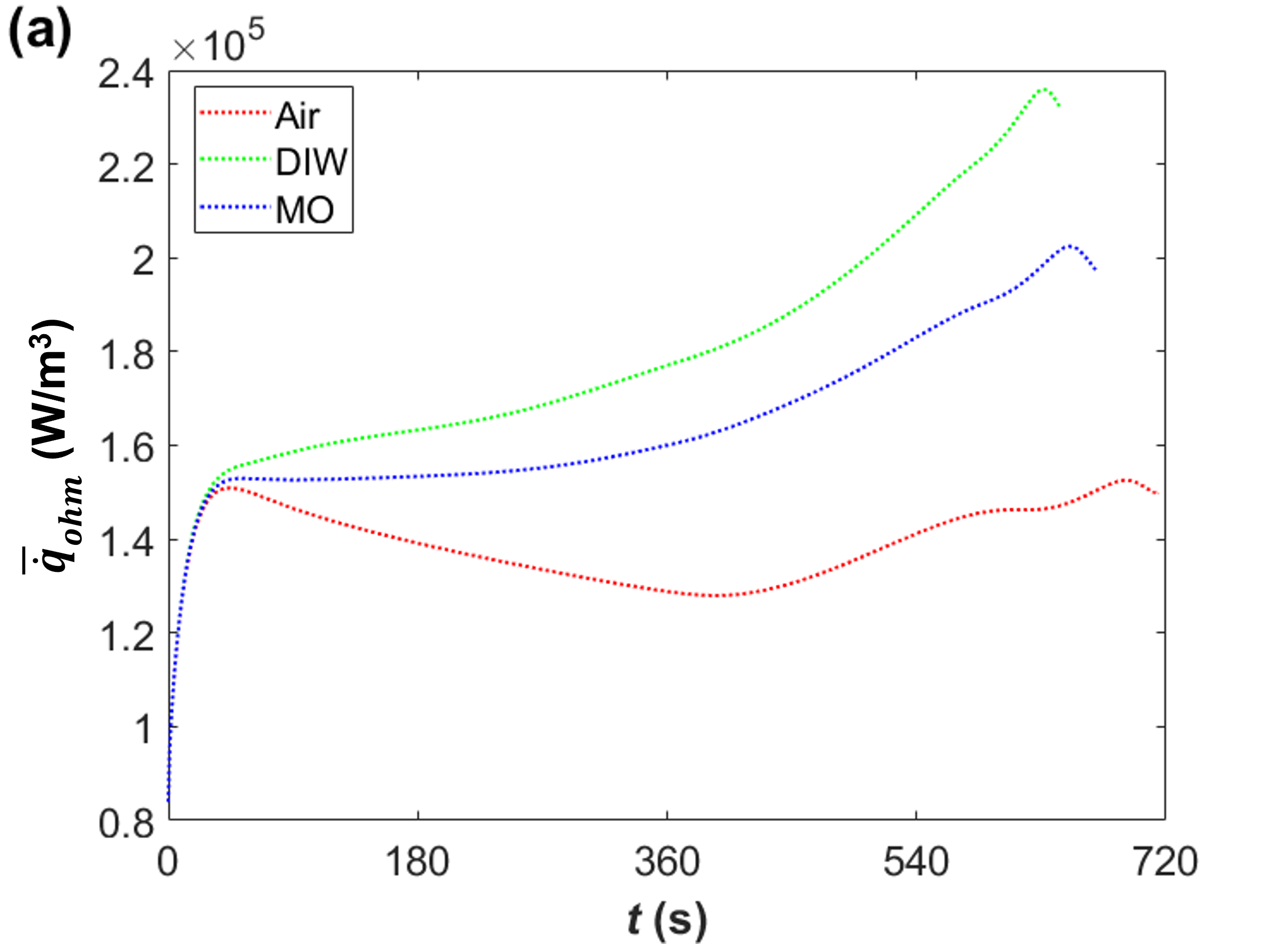}
         %\caption{}
         %\label{Fig. 10(a)}
     \end{subfigure}
     \hfill
     \begin{subfigure}[b]{0.49\textwidth}
         \centering
         \includegraphics[width=\textwidth]{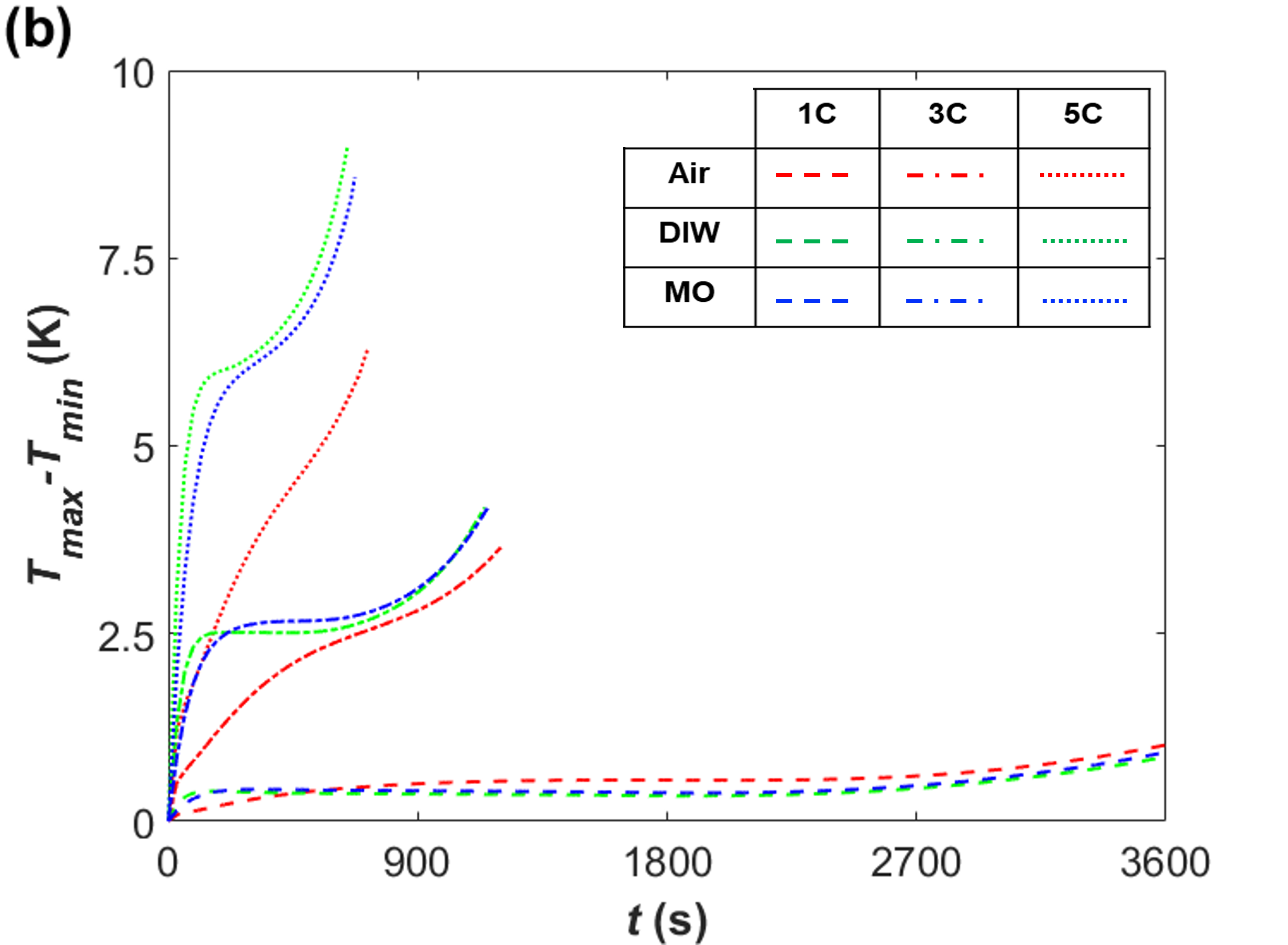}
         %\caption{}
        %\label{Fig. 10(b)}
     \end{subfigure}
     \hfill
        \caption{ (a) Average Ohmic heat generation, ($\bar{\dot{q}}_{ohm}$) for each dielectric fluids at 5C discharge rate. (b) Range of temperatures ($T_{max}$-$T_{min}$) within the battery cell for different combinations of cell discharge rate (1C, 3C, or 5C) and type of fluid (Air, DIW, or MO). Note that the mass flow rate is 0.01 kg/s for DIW \& MO and the inlet velocity is 0.3 m/s for air.}
        \label{fig: Temp difference}
\end{figure*}
As highlighted in section \ref{Effect of mass flow rate}, the DIW-based forced immersion cooling configurations have the highest average heat generation $\overline{\Dot{q}}$ despite having the lowest $\overline{\Delta T}$ rise compared to all the fluids, which is attributed to the cross-coupled nature of LIB system: the electrical resistance of the cell decreases with increasing temperature leading to lower heat generation for the poorly cooled systems. 

To gain more insight into the effect of the type of fluid on the temperature within the LIB cell, in this section, we consider the range of temperatures within the cell ($T_{max}-T_{min}$) for various combinations of discharge rate (1C, 3C, or 5C) and type of fluid (see Figure~\ref{fig: Temp difference}(b)). The range temperatures within a cell is the largest for DIW and smallest for air. This correlates with the basic heat transfer analysis, which indicates that the fluid cooling performance, as expressed by the heat transfer coefficient $h$, impacts both the average temperature of the cell and the temperature gradients within the cell as indicated by the Biot number ($Bi = h~L/\lambda_f$), which is roughly a ratio internal to external temperature differences for a system. It is used to determine when internal temperature gradients are negligible: $Bi<<1$. Fluids with effective cooling performance will have large $h$ values making it more likely that the internal gradients are significant. However, this explanation alone does not fully explain the results at a 1C discharge rate, where air has a slightly higher range of temperatures in the cell compared to the dielectric fluids. At this discharge rate, the heat generation rate is low and is fairly insensitive to the cooling fluid. Overall, these trends highlight the complex and coupled nature of heat transfer and electrochemical performance of immersion-cooled battery systems as the temperature gradient within the system is affected by the effectiveness of the external cooling system and the absolute magnitude of the temperature rise in the domain.

\section*{Appendix B: Effect of forced immersion cooling with tab cooling}\label{TabCooling}
\begin{figure*}
    \centering
    \includegraphics[scale=0.69]{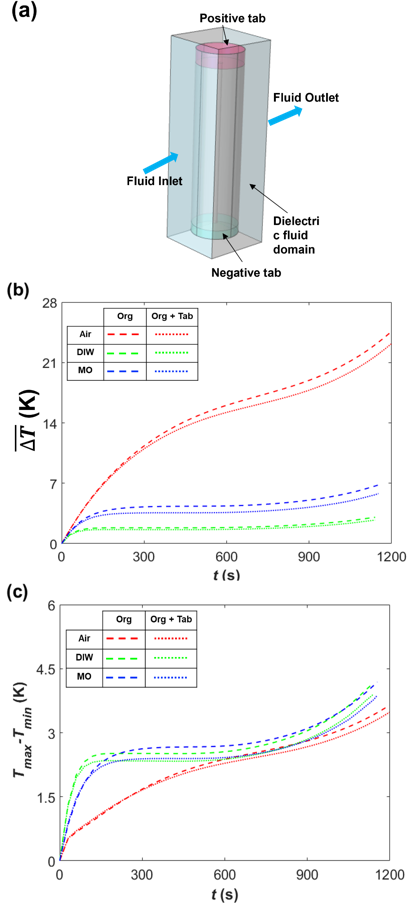}
    \caption{(a) 3D view of the fully immersed immersion cooling where fluid flows over both the positive and negative tabs, as well as the cylindrical surface of the cell. The flow passage height  above and below the battery  cellis $0.2~D$. All other geometrical details are the same as shown in Figure~\ref{fig: Geometry schematic comb}.
    (b) Average temperature rise, $\overline{\Delta T}$, and (C) range of temperatures, $T_{max}-T_{min}$, in the cylindrical jellyroll domain for different fluids at a 3C discharge rate for both the tab cooling (`Org +Tab') and nominal (`Org') configurations. Note that the mass flow rate of dielectric fluids is 0.01 kg/s and the air inlet velocity is fixed at 0.3 m/s.}
    \label{fig: FIC vs SIC-1 }
\end{figure*}
% \begin{figure*}
%      \centering
%      \begin{subfigure}[b]{0.31\textwidth}
%          \centering
%          \includegraphics[width=\textwidth]{Battery Model-1/geom_3D_final_FIC.png}
%          %\caption{}
%          \label{Fig. 11(a)}
%      \end{subfigure}
%      \hfill
%      \begin{subfigure}[b]{0.48\textwidth}
%          \centering
%          \includegraphics[width=\textwidth]{Battery Model-1/temp_FICvsSIC.png}
%          %\caption{}
%          \label{Fig. 11(b)}
%      \end{subfigure}
%      \hfill
%      \begin{subfigure}[b]{0.48\textwidth}
%          \centering
%          \includegraphics[width=\textwidth]{Battery Model-1/dT_FICvsSIC.png}
%          %\caption{}
%          \label{Fig. 11(c)}
%      \end{subfigure}
%      \hfill
%         \caption{(a) 3D view of the full flow immersed cooling (F-FIC) where fluid flows over the both the tabs (\textit{i.e.}, positive and negative) in the passage height of 0.2$\times D$ and rest all the geometrical details such as $\Delta X_{perp}$ and $\Delta X_{par}$ are same as FIC shown in Figure \ref{fig: Geometry schematic comb}. (b) Average temperature rise, $\overline{\Delta T}$ in the cylindrical jellyroll domain and (C) $T_{max}-T_{min}$ for different fluids at at 3C discharge rate operating under F-FIC \& FIC configurations. Note that mass flow rate of dielectric fluids (\textit{i.e.}, DIW \& MO) is 0.01 kg/s and air inlet velocity is fixed at 0.3 m/s similar to section \ref{Effect of discharge rate}.}
%         \label{fig: FIC vs SIC-1}
% \end{figure*}
For the geometry evaluated thus far (see Section \ref{Geometry}), only the lateral (cylindrical) surface of the LIB cell is cooled in the original geometry (see Figure \ref{fig: Geometry schematic comb}). However, flowing the immersion cooling fluid over the tabs, in addition to the lateral surface of the cell, can improve performance \cite{Li2021OptimalBatteries} in part because the thermal conductivity of the jellyroll is high in the axial direction making this pathway for heat removal more efficient. Figure \ref{fig: FIC vs SIC-1 } (a) shows the forced immersion cooling with tab cooling (Org +Tab) geometry, where 'Org' means the original configuration shown in Figure \ref{fig: Geometry schematic comb}. Apart from the added flow passages above and below the cell (of height $0.2~D$), all other geometrical parameters are the same as Figure \ref{fig: Geometry schematic comb}. Here, we focus on a discharge rate of 3C with a mass flow rate of 0.01 kg/s for DIW and MO. The results are compared to air with an inlet velocity of 0.3 m/s. 

Figure \ref{fig: FIC vs SIC-1 }(b) shows that the average temperature rise decreases slightly with the addition of tab cooling for cells operating at similar conditions. The change in the temperature rise is the largest for air and smallest for DIW, which is directly related to the cooling performance of the fluid. Higher cooling capacity reduces the average temperature rise. Therefore, adding tab cooling has a smaller effect for these systems than the air-cooled case. Moreover, the range of the temperature distribution within the LIB cell is decreased with the addition of tab cooling as illustrated in Figure \ref{fig: FIC vs SIC-1 }(c). Note that these results are for discharging at 3C and 0.01 m/s, but the performance may vary for different combinations of discharge rate and mass flow rate similar to the response observed for the case without tab cooling.
 
\bibliographystyle{unsrtnat}
\bibliography{references}

\end{document}